\documentclass[aps, prl, tightenlines,superscriptaddress, footinbib,reprint]{revtex4-2}
\usepackage{amsmath}
\usepackage{graphicx}
\usepackage{color}
\usepackage{amsfonts}
\usepackage{txfonts}
\usepackage[colorlinks,citecolor=blue]{hyperref}

\begin{document}

\title{Manipulating Topological Polaritons in Optomechanical Ladders}
\author{Jia-Kang Wu}
\affiliation{Key Laboratory of Low-Dimensional Quantum Structures and Quantum Control of
Ministry of Education, Key Laboratory for Matter Microstructure and Function of Hunan Province, Department of Physics and Synergetic Innovation Center for Quantum Effects and Applications, Hunan Normal University, Changsha 410081, China}

\author{Xun-Wei Xu}
\email{Corresponding author: xwxu@hunnu.edu.cn}
\affiliation{Key Laboratory of Low-Dimensional Quantum Structures and Quantum Control of
Ministry of Education, Key Laboratory for Matter Microstructure and Function of Hunan Province, Department of Physics and Synergetic Innovation Center for Quantum Effects and Applications, Hunan Normal University, Changsha 410081, China}
\affiliation{Institute of Interdisciplinary Studies, Hunan Normal University, Changsha, 410081, China}

\author{Hui Jing}
\affiliation{Key Laboratory of Low-Dimensional Quantum Structures and Quantum Control of
Ministry of Education, Key Laboratory for Matter Microstructure and Function of Hunan Province, Department of Physics and Synergetic Innovation Center for Quantum Effects and Applications, Hunan Normal University, Changsha 410081, China}

\author{Le-Man Kuang}
\affiliation{Key Laboratory of Low-Dimensional Quantum Structures and Quantum Control of
Ministry of Education, Key Laboratory for Matter Microstructure and Function of Hunan Province, Department of Physics and Synergetic Innovation Center for Quantum Effects and Applications, Hunan Normal University, Changsha 410081, China}

\author{Franco Nori}
\affiliation{Theoretical Quantum Physics Laboratory, Cluster for Pioneering Research, RIKEN, Wakoshi, Saitama, 351-0198, Japan}
\affiliation{Quantum Computing Center, RIKEN, Wakoshi, Saitama, 351-0198, Japan}
\affiliation{Physics Department, University of Michigan, Ann Arbor, MI 48109-1040, USA}

\author{Jie-Qiao Liao}
\email{Corresponding author: jqliao@hunnu.edu.cn}
\affiliation{Key Laboratory of Low-Dimensional Quantum Structures and Quantum Control of
Ministry of Education, Key Laboratory for Matter Microstructure and Function of Hunan Province, Department of Physics and Synergetic Innovation Center for Quantum Effects and Applications, Hunan Normal University, Changsha 410081, China}
\affiliation{Institute of Interdisciplinary Studies, Hunan Normal University, Changsha, 410081, China}

\date{\today}

\begin{abstract}
We propose to manipulate topological polaritons in optomechanical ladders consisting of an optical Su-Schrieffer-Heeger (SSH) chain and a mechanical SSH chain connected through optomechanical (interchain) interactions.
We show that the topological phase diagrams are divided into six areas by four boundaries and that there are four topological phases characterized by the Berry phases.
We find that a topologically nontrivial phase of the polaritons is generated by the optomechanical interaction between the optical and mechanical SSH chains even though they are both in the topologically trivial phases.
Counter-intuitively, six edge states appear in one of the topological phases with only two topological nontrivial bands,
and some edge states are localized near but not at the boundaries of an open-boundary ladder.
Moreover, a two-dimensional Chern insulator with higher Chern numbers is simulated by introducing proper periodical adiabatic modulations of the driving amplitude and frequency.
Our work not only opens a route towards topological polaritons manipulation by optomachanical interactions, but also will exert a far-reaching influence on designing topologically protected polaritonic devices.
\end{abstract}

\maketitle
\narrowtext

\emph{{\color{blue}Introduction}.---}Optomechanical couplings lie at the heart of cavity optomechanics~\cite{Aspelmeyer2014RMP,Bowen2015CRC}, and provide the physical origin for studying both fundamental physics~\cite{Schwab2005PhT,Abbott2016PRL} and modern quantum technologies~\cite{Barzanjeh2022NatPh}.
An impressive series of milestones have been achieved in optomechanical systems with single- or few-mode cavities, such as ground-state cooling of nanomechanical resonators~\cite{Wilson-Rae2007PRL,Marquardt2007PRL,Teufel2011Natur,Chan2011Natur}, normal mode splitting~\cite{Dobrindt2008PRL,Grblacher2009Nat,Teufel2011Nature,Verhagen2012Nature}, optomechanical correlations and entanglement~\cite{Vitali2007PRL,Riedinger2018Natur,Ockeloen2018Natur,YuHC2020Natur,Kotler2021Sci,Mercier2021Sci}, quantum squeezing of mechanical motions~\cite{Wollman2015Sci,Pirkkalainen2015PRL,Lecocq2015PRX,Lei2016PRL}, and position measurements at the quantum level~\cite{Metcalfe2014APR,Schliesser2009NatPh,Teufel2009NatNa}.
Recent advances in fabrication, manipulation, and detection of optomechanical systems pave the way to exploring many-body physics in optomechanical arrays, mainly focusing on
synchronization of many mechanical resonators~\cite{Heinrich2011PRL,Holmes2012PRE,Ludwig2013PRL,Lauter2015PRE,Zhang2015PRL,ZhengX2021PRA}, Dirac and gauge physics~\cite{Schmidt2015NJP,Schmidt2015Optic,Seif2018NatCo,Mathew2020NatNa,Denis2020PRL}, various localization behaviors~\cite{Xuereb2014PRL,Figueiredo2017NJPh,XiongH2017PRL,Lemonde2019NJPh}, and optomechanical topology for photons and phonons~\cite{Peano2015PRX,Sanavio2020PRB,Shah2022arXiv,QiLu2017OExpr,Lemonde2019NJPh,Brendel2018PRB,XuXW2022FrP,Raeisi2020PRA,Ni2021OPTICA,HaoXZ2022PRA}. In addition, topological optomechanical lattices described by the Su-Schrieffer-Heeger (SSH)~\cite{Su1979PRL,Janos2016Springer,Ozawa2019RMP,CheY2020PRB}, Kitaev~\cite{slim2024Nat,WanLL2023PRL}, and graphene~\cite{Castro2009RMP,Hasan2010RMP,QiXL2011RMP,Bernevig2013,Naumis_2017} models have been recently realized in optomechanical crystals~\cite{RenHJ2022NatCo,slim2024Nat} and superconducting circuit optomechanics~\cite{Youssefi2021arXiv}.

Topological optomechanics in the strong (linearized) optomechanical coupling regime~\cite{Dobrindt2008PRL,Grblacher2009Nat,Teufel2011Nature,Verhagen2012Nature} should be considered with an insight from the polaritons of quasi-particles formed by a strong mixing of photons and phonons~\cite{Ranfagni2021NatPh,Hughes2021PRB}.
That is clearly different from topological optomechanical lattices for investigating topological phononics~\cite{RenHJ2022NatCo} or topological microwave modes~\cite{Youssefi2021arXiv}.
Polaritons~\cite{Huang1951Natur,Tolpygo1950,Hopfield1958} exhibit a dual nature of light and matter, and the topological properties of polaritons can be controlled by light-matter interactions~\cite{Bardyn2015PRB,Karzig2015PRX,HuG2020Natur,Guddala2021Sci,LiM2022SciA,HuH2022NatNa}.
In particular, the optomechanical interactions are tunable by external optical pumping~\cite{Dobrindt2008PRL,Grblacher2009Nat,Teufel2011Nature,Verhagen2012Nature}, which provides wide opportunities for manipulating topological optomechanical polaritons artificially. Moreover, optomechanical lattices with a variety of topological phases provide an ideal platform for exploring exotic quantum light-matter interactions~\cite{Perczel2017PRL,Barik2018Sci,DongXL2021PRL}.

In this Letter, we propose to manipulate the topological states of polaritons in an optomechanical ladder consisting of an optical and a mechanical SSH chains coupled through optomechanical interactions.
We show that the \emph{topological states of the polaritons in an optomechanical ladder can be tuned on demand} by adjusting the amplitude of the driving fields.
Moreover, a two-dimensional (2D) \emph{Chern insulator for polaritons} is demonstrated in the optomechanical ladder by adiabatically and periodically modulating the amplitude and frequency of the driving fields.
Our proposal not only opens a route towards \emph{topological insulators of optomachanical polaritons}, but also inspires potential applications in designing topologically protected quantum devices~\cite{Harari2018SCI,Bandres2018SCI,Andrea2018Sci,NieW2020PRL,NieW2021PRL,Mehrabad2023arXiv}.

\begin{figure}[tbp]
\includegraphics[bb=154 453 439 644, width=8.5cm, clip]{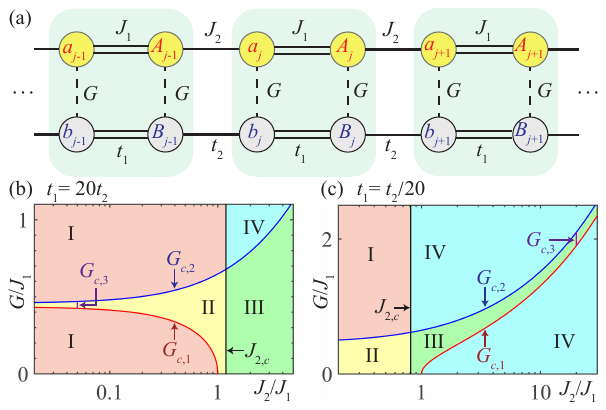}
\caption{(Color online) (a) Schematic of an optomechanical ladder. Two SSH chains of optical modes ($a_{j}$ and $A_{j}$) and mechanical modes ($b_{j}$ and $B_{j}$), shown by yellow and gray circles with staggered mutual couplings $\{J_{1},J_{2}\}$ and $\{t_{1}, t_{2}\}$, are coupled by the linearized optomechanical interaction with strength $G$. Topological phase diagram ($\delta=0$) at different parameters: (b) $\{t_1=J_1/5,t_2=J_1/100\}$ and (c) $\{t_1=J_1/100,t_2=J_1/5\}$.}
\label{Fig1}
\end{figure}

\emph{{\color{blue}Optomechanical ladders and topological phases}.---}We consider an optomechanical
ladder formed by an optical and a mechanical SSH chains coupled through linearized optomechanical interactions (see Supplemental Material~\cite{SM} for details) [Fig.~\ref{Fig1}(a)]. In the strong driving regime of all the optical modes, the linearized Hamiltonian of this optomechanical ladder reads ($\hbar =1$)
\begin{eqnarray}
H_{\mathrm{sys}} &=& \sum_{j}\left[\delta( a_{j}^{\dag }a_{j}-A_{j}^{\dag}A_{j}) -G(a_{j}^{\dag }b_{j}+A_{j}^{\dag }B_{j}+\mathrm{H.c.})\right.\nonumber\\
&+ &\left. ( J_{1}A_{j}^{\dag }a_{j}+J_{2}a_{j+1}^{\dag
}A_{j}+t_{1}B_{j}^{\dag }b_{j}+t_{2}b_{j+1}^{\dag }B_{j}+\mathrm{H.c.})\right],
\end{eqnarray}
where $a_{j}$ and $A_{j}$ ($b_{j}$ and $B_{j}$) are annihilation operators of
the optical (mechanical) modes at the $j$th cell.
The parameter $\delta=\Delta_a-\omega_m=\omega_m-\Delta_A$, where $\omega_m$ is the resonance frequency of all the mechanical modes, $\Delta_a=\omega_a-\omega_d$ and $\Delta_A=\omega_A-\omega_d$ are the frequency detunings between the optical modes ($\omega_a$ for mode $a_{j}$ and $\omega_A$ for mode $A_{j}$) and the pumping fields ($\omega_d$).
$G$ is the linearized optomechanical coupling strength, which can be tuned continuously by the driving fields, and without loss of generality, hereafter we consider a real $G$ for simplicity by choosing proper driving phases; $J_{1}$ and $t_{1}$ ($J_2$ and $t_2$) are the amplitudes of intracellular (intercellular) photon- and phonon-hopping rates, respectively.

By imposing periodic boundary conditions with $N$ unit cells and introducing the Fourier transformation
$O_k=(1/\sqrt{N})\sum_{j}e^{ijkd_{0}}O_{j}$, for $O=a,\,b,\,A$, and $B$ ($k$ is the wave number and $d_0$ is the lattice constant, hereafter we set $d_0=1$ for simplicity),
the Hamiltonian in the momentum space is given by $H_{\mathrm{sys}}=\sum_{k} V_{k}^{\dagger}H_{k}V_{k}$, with $(V_{k})^{\dagger}=(a_{k}^{\dagger},A_{k}^{\dagger},b_{k}^{\dagger},B_{k}^{\dagger})$ and
\begin{equation}\label{Eq2}
H_{k}=\left(
\begin{array}{llll}
\delta  & J_{1}+J_{2}e^{ik} & -G & 0 \\
J_{1}+J_{2}e^{-ik} & -\delta  & 0 & -G \\
-G & 0 & 0 & t_{1}+t_{2}e^{ik} \\
0 & -G & t_{1}+t_{2}e^{-ik} & 0%
\end{array}%
\right).
\end{equation}%
The dispersion relations (energy bands) can be obtained by diagonalizing $H_k$, and then the boundaries for different topological phases can be analyzed via the closing of energy band gaps~\cite{SM}.
Here, the topological phase diagrams are divided into six areas by four boundaries [Figs.~\ref{Fig1}(b,\,c)]:
(i) $G_{c,1}=\sqrt{\left( J_{1}-J_{2}\right) \left(
t_{1}-t_{2}\right)}$; (ii) $G_{c,2}=\sqrt{\left( J_{1}+J_{2}\right) \left(
t_{1}+t_{2}\right)}$; (iii) $G_{c,3}=\sqrt{ J_{1}t_{1}+J_{2}t_{2} + \left(
J_{2}t_{1}+J_{1}t_{2}\right)\cos k}$ for $0<k<\pi$ and $J_{2}/J_{1}=t_{1}/t_{2}$; and (iv) $J_{2,c}=J_{1}+t_{1}-t_{2}$.
Note that three of the boundaries [i.e., (i)-(iii)] have been obtained in Refs.~\cite{Wakatsuki14PRB,Padavic2018PRB,Schnyder08PRB}, but the boundary (iv) is missed and hence only three different phases are shown there.

The topological phase of the polaritons in the optomechanical ladder can be characterized by the Berry phase set $\Lambda=\{\gamma_1,\gamma_2,\gamma_3,\gamma_4\}$ with
$\gamma_n\equiv \int^{2 \pi}_{0} A^{(n)}_k dk$, where the Berry connection $A^{(n)}_k\equiv\langle\Psi_n |i\partial_k | \Psi_n\rangle$ depends on the eigenstate $| \Psi_n\rangle$ of the Hamiltonian $H_k$ in Eq.~(\ref{Eq2}) for the $n$th band $(n=1,2,3,4)$.
In the cases of $\delta=0$, the Berry phase $\gamma_n$ takes $0$ or $\pm\pi$, corresponding to topologically trivial and nontrivial phases, respectively.
Since the ratio $t_1/t_2$ ($J_1/J_2$) plays a critical role in characterizing the phases in the SSH model~\cite{Janos2016Springer}, below we consider the cases $t_1>t_2$ and $t_1<t_2$, and study the dependence of the topological phases on the optomechanical-coupling strength $G$ and the ratio $J_1/J_2$.
Note that the arrangement for the continuous change of $t_1/t_2$ in the two cases $J_1>J_2$ and $J_1<J_2$ leads to the same physical results.

\begin{figure*}[tbp]
\includegraphics[bb=22 253 573 511, width=18cm, clip]{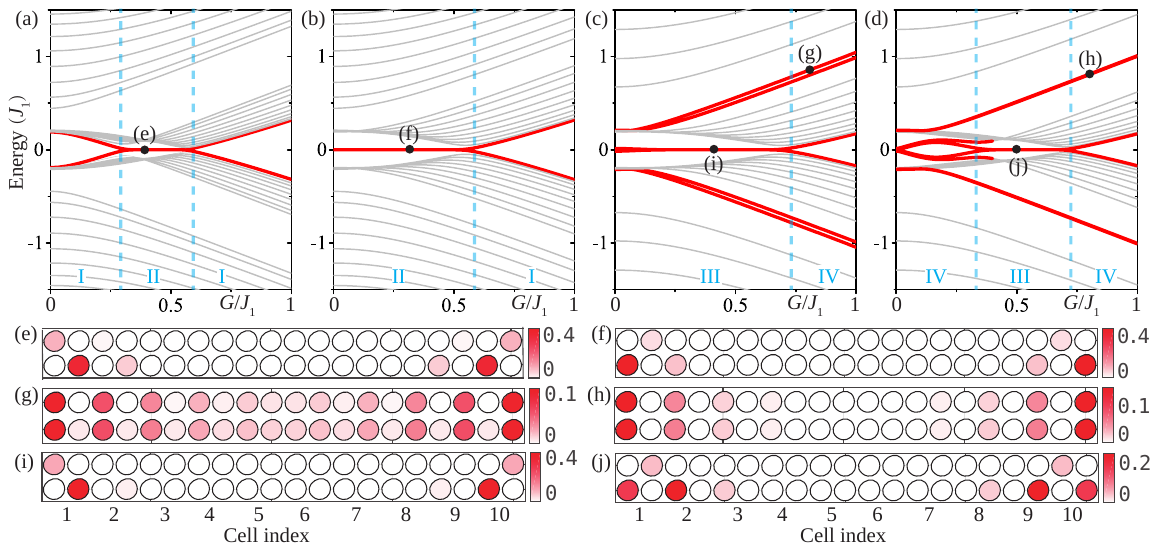}
\caption{(Color online) The energy spectrum of the system in the open-boundary condition ($N=10$) versus the optomechanical coupling $G$ for: (a) $J_2=0.6J_1$, $t_1=0.2J_1$, and $t_2=0.01J_1$, (b) $J_2=0.6J_1$, $t_1=0.01J_1$, and $t_2=0.2J_1$, (c)  $J_2=1.5J_1$, $t_1=0.2J_1$, and $t_2=0.01J_1$, (d) $J_2=1.5J_1$, $t_1=0.01J_1$, and $t_2=0.2J_1$. (e)-(j) The field distribution of the edge states marked by black dots in (a)-(d) for: (e) $G=0.4J_1$, (f) $G=0.3J_1$, (g) $G=0.8J_1$, (h) $G=0.8J_1$, (i) $G=0.4J_1$, and (j) $G=0.5J_1$.}
\label{Fig2}
\end{figure*}

There are four different phases of polaritons based on the Berry phases: I for $\Lambda=\{0,0,0,0\}$, II for $\Lambda=\{0,\pm \pi,\pm \pi,0\}$, III for $\Lambda=\{\pm \pi,0,0,\pm \pi\}$, and IV for $\Lambda=\{\pm \pi,\pm \pi,\pm \pi,\pm \pi\}$.
It is worth noting that the boundary of $G_{c,3}$ cannot be distinguished by Berry phases because the sign of the Berry phases $\pm \pi$ is meaningless.
This boundary can be verified when two optomechanical ladders in different phases are attached~\cite{SM}.

Interestingly, the optomechanical coupling $G$ connects the two SSH chains and it provides a mean to \emph{manipulate} the topological polaritons in the optomechanical ladder.
In the case $t_1>t_2$ ($t_1<t_2$) with $\delta t= |t_1-t_2|$, when $0<J_2/J_1<1$ ($1<J_2/J_1$), the system experiences phase transitions $\textrm{I}\rightarrow \textrm{II} \rightarrow \textrm{I}$ ($\textrm{IV}\rightarrow \textrm{III} \rightarrow \textrm{IV}$) with increasing $G/J_1$;
when $1<J_2/J_1<1+\delta t/J_1$ ($1-\delta t/J_1<J_2/J_1<1$), the phase transition $\textrm{II} \rightarrow \textrm{I}$ ($\textrm{III} \rightarrow \textrm{IV}$) takes place with increasing $G/J_1$; in addition, in the range $J_2/J_1>1+\delta t/J_1$ ($0<J_2/J_1<1-\delta t/J_1$), the system transits from phases III (II) to IV (I) as $G/J_1$ grows.
Note that the phase transitions \emph{only} take place \emph{either} between phases I and II \emph{or} between phases III and IV when changing $G$,
because the phases $\{\textrm{I},\textrm{II}\}$ are divided from phases $\{ \textrm{III},\textrm{IV} \}$ by the $G$-independent line $J_2/J_1=1+(t_1-t_2)/J_1$.

\emph{{\color{blue}Energy spectra and edge states}.---}According to the bulk-boundary correspondence, topological phase transitions in the optomechanical ladders can be demonstrated by the edge states in the energy spectra with an open boundary condition, and the edge states can be observed by the reflection spectra of a waveguide side-coupled to the system~\cite{Zhou2008PRL,Zhou2008PRA,Zhou2009PRA,Liao2010PRA,SM}.
To this end, the energy spectra versus $G/J_1$ are shown in Figs.~\ref{Fig2}(a-d), corresponding to the phase transitions $\textrm{I} \rightarrow \textrm{II} \rightarrow \textrm{I}$ and $\textrm{III} \rightarrow \textrm{IV}$ when $t_1>t_2$, as well as $\textrm{II} \rightarrow \textrm{I}$ and $\textrm{IV}\rightarrow \textrm{III} \rightarrow \textrm{IV}$ when $t_2>t_1$.

There are no edge states in phase I but two edge states around $E=0$ in phase II [Figs.~\ref{Fig2}(a,\,b)].
The edge states in Fig.~\ref{Fig2}(b) originate from the topologically nontrivial mechanical chain, while in Fig.~\ref{Fig2}(a)\emph{ the edge states are induced by the optomechanical interaction between the two topologically trivial chains}.
Besides, the corresponding field distributions for the edge states are different [Figs.~\ref{Fig2}(e,\,f)].
The maxima of the fields are located at the two ends of the mechanical chain for the edge state in panel~(f), while the maxima in panel (e) appear at the two modes next to the two ends of the mechanical chain.
There are four edge states around the energy $E\approx \pm G$ in phase~IV [Fig.~\ref{Fig2}(d)], which originate from the \emph{optomechanical interaction} between the two topologically nontrivial chains with frequency splittings $\pm G$.
In addition, the optomechanical interaction also induces the transition from phase III to IV [Fig.~\ref{Fig2}(c)].
The corresponding edge states are shown in Figs.~\ref{Fig2}(g,\,h), which indicate that the edge state in panel (h) is more local than the one in panel (g).

It is counter-intuitive that there are \emph{six edge states} in phase III: four appear around $E\approx \pm G$ for $G>t_1+t_2$, and the other two arise around $E=0$.
The edge states can be understood by their field distributions.
The field distributions for the edge states around $E\approx \pm G$ are very similar to the ones in the phase IV [Figs.~\ref{Fig2}(g,\,h)], which originate from the interactions between the optical edge states and the mechanical modes with frequency splitting $\pm G$.
The field distribution of the edge states around $E=0$ are shown in Figs.~\ref{Fig2}(i,\,j). Different from the traditional edge states localized at both ends, the maxima of the fields in panel (i) [(j)] appear at the first (second) two modes next to both ends of the mechanical chain.

\begin{figure*}[tbp]
\centering
\includegraphics[bb=33 411 483 647, width=18cm, clip]{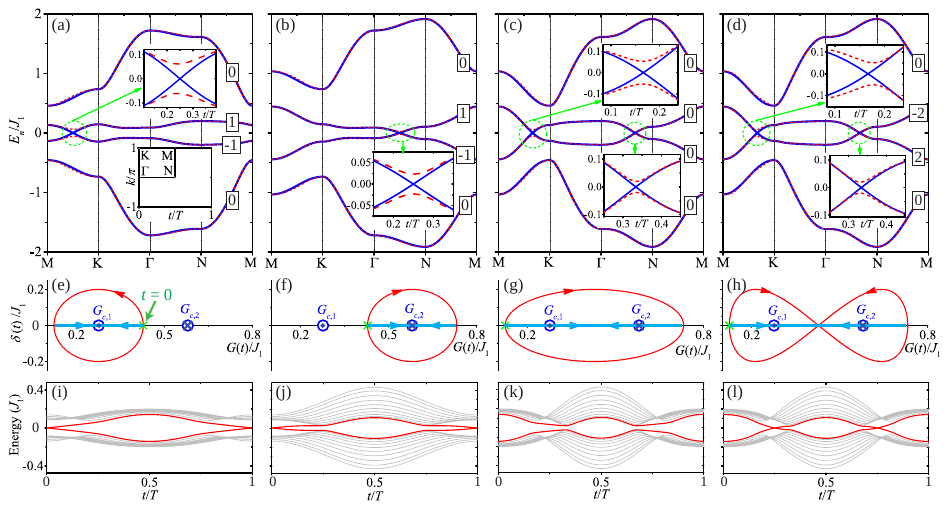}
\caption{(Color online) (a)-(d) Band diagram with quasi-momentum $k$ and time $t$ varying around the irreducible Brillouin zone of the 2D Chern insulator. The blue solid curves are plotted with $\delta=0$, and the red dashed curves are plotted with $\delta/J_1=0.2\sin(2\pi t/T')$. The Chern numbers of the bands in (a)-(d) are shown in the boxes. The band diagram in (a)-(d) and the energy spectrum for the second and third bands of $N=10$ cells in (i)-(l) are plotted based on the paths of the endpoint of the vector $\textbf{d}(t)$ shown in (e)-(h) with initial points denoted by green ``$\times$", where (e) $\bar{G}=G_{c,1}$, $\tilde{G}=(G_{c,2}-G_{c,1})/2$, $\tilde{\delta}/J_1=0.2$ (red) [$\tilde{\delta}=0$ (blue)], and $T=T'$; (f) $\bar{G}=G_{c,2}$, $\tilde{G}=(G_{c,1}-G_{c,2})/2$, $\tilde{\delta}/J_1=0.2$ (red) [$\tilde{\delta}=0$ (blue)], and $T=T'$; (g) $\bar{G}=(G_{c,1}+G_{c,2})/2$, $\tilde{G}=G_{c,1}-G_{c,2}$, $\tilde{\delta}/J_1=0.2$ (red) [$\tilde{\delta}=0$ (blue)], and $T=T'$; (h) $\bar{G}=(G_{c,1}+G_{c,2})/2$, $\tilde{G}=G_{c,1}-G_{c,2}$, $\tilde{\delta}/J_1=0.2$ (red) [$\tilde{\delta}=0$ (blue)], and $T=2T'$.}
\label{Fig3}
\end{figure*}

\emph{{\color{blue}2D Chern insulators by adiabatic modulation}.---}The optomechanical ladders can simulate a 2D Chern insulator with the wave number for the second dimension replaced by the time dimension in adiabatical modulation, which can be realized by tuning the driving strength and frequency periodically.
The energy band diagrams for time-dependent strength $G(t)=\bar{G}+\tilde{G}\cos(2 \pi t/T)$ are shown by the blue solid curves in Figs.~\ref{Fig3}(a)-\ref{Fig3}(d), with the first Brillouin zone in the inset of Fig.~\ref{Fig3}(a), where $\bar{G}$ and $\tilde{G}$ are real numbers and $T$ is the modulation period.
There are two kinds of Dirac points in the bands located at $\{k=\pm \pi,\,G(t)=G_{c,1}\}$ along the line $\rm MK$ and at $\{k=0,\,G(t)=G_{c,2}\}$ along the line $\Gamma {\rm N}$.
To open a gap around the Dirac points, we introduce a time-dependent detuning $\delta(t)=\tilde{\delta}\sin(2 \pi t/T')$ by modulating the frequency with amplitude $\tilde{\delta}$ and period $T'$.
To be more intuitive, we define a vector $ \textbf{d}(t)\equiv G(t) \textbf{d}_x+ \delta(t) \textbf{d}_y$,
with unit vectors $\textbf{d}_x$ and $\textbf{d}_y$ in the $x$ and $y$ directions.
Then the path of the endpoint of $\textbf{d}(t)$ is a closed loop due to the periodicity of the parameters $G(t)$ and $\delta(t)$.
In Figs.~\ref{Fig3}(e-h), we show four different modulation schemes for the loop winding around either one or two of the critical points $G_{c,1}$ and $G_{c,2}$.
Note that the red trajectories of the parameters $G(t)$ and $\delta(t)$ in Figs.~\ref{Fig3}(e)-\ref{Fig3}(h) correspond to the red dashed energy bands in Figs.~\ref{Fig3}(a)-\ref{Fig3}(d), while the blue trajectories of $G(t)$ (with $\delta=0$) correspond to the blue bands with the Dirac points.
These plots indicate that the Dirac point can be opened by introducing a time-dependent $\delta(t)$.

To identify the topological phase of the polaritons under adiabatic pumping, we seek for the Chern numbers~\cite{Janos2016Springer} of the $n$th band $
c_n=-\frac{1}{2\pi}\int^{2 \pi}_{0}dk\int^T_{0}dt \: ( \partial A^{(n)}_t/\partial k - \partial A^{(n)}_k/\partial t )$,
where the Berry connections $A^{(n)}_t\equiv \langle\Psi_n | i\partial_t | \Psi_n\rangle$ and $A^{(n)}_k\equiv \langle\Psi_n | i\partial_k | \Psi_n\rangle$ depend on both $k$ and $t$ via $G(t)$ and $\delta (t)$.
The Chern numbers for the energy bands are also shown in Figs.~\ref{Fig3}(a)-\ref{Fig3}(d).
When the loop winds around either the critical point $G_{c,1}$ counter-clockwise or $G_{c,2}$ clockwise, the Chern-number set of the four bands becomes $C=\{0,-1,1,0\}$.
When the loop winds around both $G_{c,1}$ and $G_{c,2}$ clockwise or counter-clockwise [Fig.~\ref{Fig3}(g)], the Chern numbers are zero for all bands.
More interestingly, when the loop is a Lissajous figure with a period ratio of $T:T'=2:1$ [Fig.~\ref{Fig3}(h)], the Chern-number set becomes $C=\{0,2, -2, 0\}$, which implies the construction of \emph{a topological model with higher Chern-number bands}, providing a promising platform for exploring new topological states~\cite{WenXG2019SCI,WangJ2017NatMa,Barkeshli2012PRX,ZhaoL2012PRL,Skirlo2015PRL}.

The Chern numbers can be confirmed by examining the number of the edges states across the bulk gap with an open boundary condition.
In Figs.~\ref{Fig3}(i)-\ref{Fig3}(l), we show the edge state branches connecting the second (lower) and third (upper) bands (the first and fourth bands are not shown in the figures for $c_1=c_4=0$).
In Figs.~\ref{Fig3}(i) and \ref{Fig3}(j), there exist both one edge state propagating from the upper to the lower bands and one edge state propagating from the lower to the upper bands, corresponding to the Chern numbers $c_2=-1$ and $c_3=1$.
There is no edge state across the bulk gap in Fig.~\ref{Fig3}(k), which is related to $c_2=c_3=0$ for both the upper and lower bands.
For higher Chern numbers $c_2=2$ and $c_3=-2$, there are both two edge states propagating from the upper to lower bands and two edge states propagating from lower to upper bands [Fig.~\ref{Fig3}(l)].
Moreover, the Chern numbers can also be verified by the adiabatic particle pumping process~\cite{SM}.

\emph{{\color{blue}Experimental feasibility}.---}The physical platform candidates for implementing the optomechanical ladders should satisfy the following conditions: (i) mechanical frequency $\omega_m \gg \{J_1,J_2,t_1,t_2,G,\delta\}$ for making the rotating-wave approximation; (ii) strong-coupling conditions: $\{J_1,J_2,t_1,t_2,G\} > \{\kappa,\gamma\}$ ($\kappa$ and $\gamma$ are the optical and mechanical damping rates, respectively).
Optomechanical crystals (OMCs)~\cite{Eichenfield2009Natur,Gomis2014NatCo} are one of the appropriate platforms that satisfy all these conditions.
As reported in the experiments~\cite{Fang2016NaPho,Fang2017NatPh,Burek2016Optic}, most of the OMCs operate in the resolved-sideband regime, with the mechanical frequency ranging from a few GHz to about $10$ GHz.
Moreover, the linearized optomechanical coupling rate $G/2\pi$ (from a few MHz to about $100$~MHz) can be controlled by the optical pump, and the photon- and phonon-hopping rates $J/2\pi \sim 500$ MHz and $t/2\pi \sim 200$ MHz can be designed as needed~\cite{Seif2018NatCo}.
With a high-quality factor around $10^7$ for both optical and mechanical modes~\cite{MacCabe2020Sci,Ren2020NatCo,Asano2017OExpr}, we have damping rates $\kappa/2\pi \sim 20$ MHz and $\gamma/2\pi \sim 1$ kHz, to ensure the system working in the strong-coupling regime.
These analyses indicate that our proposal can be realized in the state-of-the-art setups.

\emph{{\color{blue}Conclusions}.---}We have investigated the topological properties of polaritons in optomechanical ladders consisting of an optical and a mechanical SSH chains connected through optomechanical interactions.
A set of \emph{four different topological phases and the transitions between them have been explored by adjusting the amplitude of the optomechanical interactions}.
We have also shown that \emph{a 2D Chern insulator} can be implemented by adiabatically modulating the parameters in optomechanical ladders.
Our work opens a route towards exploring rich topological states for polaritons in optomechanics, which can be applied for developing topologically protected optomechanical technologies.

\begin{acknowledgments}
We thank Prof.~Xiong-Jun Liu, Prof.~Tao Liu, Prof.~Wei Nie, and Prof.~Ziming Zhu for helpful suggestions.
X.-W.X. was supported by National Natural Science Foundation of China (NSFC) (Grants No.~12064010 and  No.~12247105),
the science and technology innovation Program of Hunan Province (Grant No.~2022RC1203),
and Hunan provincial major sci-tech program (Grant No.~2023ZJ1010). J.-Q.L. was supported in part by the NSFC (Grants
No.~12175061, No.~11935006, and  No.~12247105), the Science and
Technology Innovation Program of Hunan Province (Grants No.~2021RC4029 and
No.~2020RC4047), and Hunan provincial major sci-tech program (Grant No.~2023ZJ1010).
H.J. was supported by the NSFC (Grants No.~11935006 and No.~11774086), the Science and Technology Innovation
Program of Hunan Province (Grant No.~2020RC4047), and Hunan provincial major sci-tech program (Grant No.~2023ZJ1010).
L.-M.K. was supported by the NSFC (Grants No.~1217050862, No.~11935006, No.~11775075, and No.~12247105) and the Science and Technology Innovation Program of Hunan Province (Grant No.~2020RC4047).
F.N. is supported in part by: Nippon Telegraph
and Telephone Corporation (NTT) Research, the Japan
Science and Technology Agency (JST) [via the Quantum
Leap Flagship Program (Q-LEAP), and the Moonshot
R\&D Grant Number~JPMJMS2061], the Asian Office
of Aerospace Research and Development (AOARD) (via
Grant No.~FA2386-20-1-4069), and the Foundational
Questions Institute Fund (FQXi) via Grant No.~FQXi-
IAF19-06.
\end{acknowledgments}

\bibliography{ref}


\onecolumngrid
\newpage
\setcounter{equation}{0} \setcounter{figure}{0}
\setcounter{table}{0}

\renewcommand{\theequation}{S\arabic{equation}}
\renewcommand{\thefigure}{S\arabic{figure}}
\renewcommand{\bibnumfmt}[1]{[S#1]}
\renewcommand\thesection{S\arabic{section}}

\setcounter{page}{1}\setcounter{secnumdepth}{3} \makeatletter

\begin{center}
{\large \bf Supplementary Material for ``Manipulating Topological Polaritons in Optomechanical Ladders''}
\end{center}

This supplementary material provides the detailed calculations and results on the following four topics: (S1) derivation of the linearized Hamiltonian of the optomechanical ladder, (S2) calculation of the boundary curves of the topological phases, (S3) detection of the edge states, and (S4) adiabatic optomechanical pumping.

\section{derivation of the linearized Hamiltonian of the optomechanical ladder}

In this section, we present a detailed derivation of the linearized Hamiltonian [Eq.~(1) in the main text] of the optomechanical ladder, which consists of an optical and a mechanical Su-Schrieffer-Heeger (SSH) chains coupled via optomechanical interactions.
For manipulating the optomechanical ladder, we introduce optical pumpings to all these optical modes.
The Hamiltonian of the optomechanical ladder reads~\cite{Eichenfield2009Natur,Seif2018NatCo} ($\hbar =1$)
\begin{eqnarray}\label{Eq1}
H_{\rm oml} &=&\sum_{j}(\omega _{a} a_{j}^{\dag }a_{j}+\omega _{A} A_{j}^{\dag }A_{j})+\omega
_{m}\sum_{j}(b_{j}^{\dag }b_{j}+B_{j}^{\dag }B_{j})  \notag \\
&&-g_{1}\sum_{j}a_{j}^{\dag }a_{j}(b_{j}^{\dag
}+b_{j})-g_{2}\sum_{j}A_{j}^{\dag }A_{j}(B_{j}^{\dag }+B_{j})  \notag \\
&&+J_{1}\sum_{j}(A_{j}^{\dag }a_{j}+a_{j}^{\dag
}A_{j})+J_{2}\sum_{j}(A_{j}^{\dag }a_{j+1}+a_{j+1}^{\dag }A_{j})  \notag \\
&&+t_{1}\sum_{j}(B_{j}^{\dag }b_{j}+b_{j}^{\dag
}B_{j})+t_{2}\sum_{j}(B_{j}^{\dag }b_{j+1}+b_{j+1}^{\dag }B_{j})  \notag \\
&&+\sum_{j}(\varepsilon _{j}a_{j}^{\dag }e^{i\omega _{d}t}+\varepsilon
_{j}^{\ast }a_{j}e^{-i\omega _{d}t})+\sum_{j}(\epsilon _{j}A_{j}^{\dag
}e^{i\omega _{d}t}+\epsilon _{j}^{\ast }A_{j}e^{-i\omega _{d}t}),
\end{eqnarray}%
where $a_{j}$ ($A_{j}$) and $b_{j}$ ($B_{j}$) are the annihilation operators of the optical and mechanical modes at the $j$th site with resonance frequencies $\omega _{a}$ ($\omega _{A}$) and $\omega _{m}$ ($\omega _{m}$), respectively. The parameters $g_{1}$ and $g_{2}$ are the single-photon optomechanical interaction strengths, $J_{1}$ and $J_{2}$ ($t_{1}$ and $t_{2}$) are the tunneling strengths between the optical (mechanical) modes, and $\varepsilon _{j}$ and $\epsilon _{j}$ are the pumping amplitudes of the optical modes with the same frequency $\omega_d$. In a rotating frame defined by the unitary transformation operator $U_1=\exp[-i(\omega_{d}t)\sum_{j}(a_{j}^{\dag }a_{j}+A_{j}^{\dag }A_{j})]$, the Hamiltonian (\ref{Eq1}) becomes%
\begin{eqnarray}
H'_{\rm oml} &=& \sum_{j}(\Delta_a a_{j}^{\dag }a_{j}+ \Delta_A A_{j}^{\dag
}A_{j})+\omega _{m}\sum_{j}(b_{j}^{\dag }b_{j}+B_{j}^{\dag }B_{j})  \notag \\
&&-g_{1}\sum_{j}a_{j}^{\dag }a_{j}(b_{j}^{\dag
}+b_{j})-g_{2}\sum_{n}A_{j}^{\dag }A_{j}(B_{j}^{\dag }+B_{j})  \notag \\
&&+J_{1}\sum_{j}(A_{j}^{\dag }a_{j}+a_{j}^{\dag
}A_{j})+J_{2}\sum_{j}(A_{j}^{\dag }a_{j+1}+a_{j+1}^{\dag }A_{j}) \notag \\
&&+t_{1}\sum_{j}(B_{j}^{\dag }b_{j}+b_{j}^{\dag
}B_{j})+t_{2}\sum_{j}(B_{j}^{\dag }b_{j+1}+b_{j+1}^{\dag }B_{j})  \notag \\
&&+\sum_{j}(\varepsilon_j a_{j}^{\dag }+\varepsilon_j ^{\ast
}a_{j})+\sum_{j}(\epsilon_j A_{j}^{\dag }+\epsilon_j ^{\ast }A_{j}),
\end{eqnarray}%
where $\Delta_a =\omega _{a}-\omega _{d}$ and $\Delta_A =\omega _{A}-\omega _{d}$ are the frequency detunings between the optical modes and the pumping fields.

In this work, we assume that the \emph{external pumping fields are strong} (i.e., $|\varepsilon_j|\gg\kappa$ and $|\epsilon_j|\gg\kappa$, $\kappa$ is the damping rate of the optical modes). Then, we can apply the standard linearization procedure to obtain the linearized optomechanical Hamiltonian. By adding both the dissipation and noise terms into the Heisenberg equations, the quantum Langevin equations of this system can be obtained as%
\begin{subequations}
\label{L-H}
\begin{align}
\dot{a}_{j}=&-(i\Delta_a +\kappa )a_{j}+ig_{1}a_{j}(b_{j}^{\dag
}+b_{j})-iJ_{1}A_{j}-iJ_{2}A_{j-1}-i\varepsilon _{j}+ \sqrt {2\kappa} a_{j}^{\rm in},\\
\dot{A}_{j}=&-(i\Delta_A +\kappa )A_{j}+ig_{2}A_{j}(B_{j}^{\dag
}+B_{j})-iJ_{1}a_{j}-iJ_{2}a_{j+1}-i\epsilon _{j}+ \sqrt {2\kappa} A_{j}^{\rm in},\\
\dot{b}_{j}=&-(i\omega _{m}+\gamma )b_{j}+ig_{1}a_{j}^{\dag
}a_{j}-it_{1}B_{j}-it_{2}B_{j-1}+ \sqrt {2\gamma} b_{j}^{\rm in},\\
\dot{B}_{j}=&-(i\omega _{m}+\gamma )B_{j}+ig_{2}A_{j}^{\dag
}A_{j}-it_{1}b_{j}-it_{2}b_{j+1} + \sqrt {2\gamma} B_{j}^{\rm in},
\end{align}
\end{subequations}
where $\gamma$ is the damping rate of the mechanical modes, and $a_{j}^{\rm in}$, $A_{j}^{\rm in}$, $b_{j}^{\rm in}$, and $B_{j}^{\rm in}$ are the input noise operators of these optical and mechanical modes.

In the \emph{strong-driving regime}, the optical modes will be largely displaced; then we can linearize the optomechanical interactions around the steady state. Concretely, we separate the classical displacement and the quantum fluctuation by expressing these operators as the sum of their steady-state mean values and quantum fluctuations, namely,
\begin{subequations}
\begin{align}
a_{j}\rightarrow a_{j}+\alpha _{j},\\
A_{j}\rightarrow A_{j}+\zeta _{j},\\
b_{j}\rightarrow b_{j}+\beta _{j},\\
B_{j}\rightarrow B_{j}+\eta _{j},
\end{align}
\end{subequations}%
where $\alpha _{j}$, $\zeta _{j}$, $\beta _{j}$, and $\eta _{j}$ are the steady-state mean values, and $a_{j}$, $A_{j}$, $b_{j}$, and $B_{j}$ on the right-hand side of the arrows are the quantum fluctuation operators.
The steady-state mean values are determined by the following equations
\begin{subequations}
\label{classical}
\begin{align}
-(i\Delta'_a +\kappa )\alpha _{j}-iJ_{1}\zeta _{j}-iJ_{2}\zeta _{j-1}-i\varepsilon
_{j}=& 0,
\\
-(i\Delta'_A +\kappa )\zeta _{j}-iJ_{1}\alpha _{j}-iJ_{2}\alpha _{j+1}-i\epsilon
_{j}=& 0,
\\
-(i\omega _{m}+\gamma )\beta _{j}+ig_{1}|\alpha _{j}|^{2}-it_{1}\eta _{j}-it_{2}\eta _{j-1}=& 0,
\\
-(i\omega _{m}+\gamma )\eta _{j}+ig_{2}|\zeta _{j}|^{2}-it_{1}\beta _{j}-it_{2}\beta _{j+1}=& 0,
\end{align}%
\end{subequations}
and the linearized equations of motion for these quantum fluctuations are given by%
\begin{subequations}
\label{fluctuations}
\begin{align}
\dot{a}_{j}=& -(i\Delta'_{a} +\kappa )a_{j}+iG_{j,1}(b_{j}^{\dag
}+b_{j})-iJ_{1}A_{j}-iJ_{2}A_{j-1}+\sqrt {2\kappa}a_{j}^{\rm in},
\\
\dot{A}_{j}=& -(i\Delta'_{A} +\kappa )A_{j}+iG_{j,2}(B_{j}^{\dag
}+B_{j})-iJ_{1}a_{j}-iJ_{2}a_{j+1}+\sqrt {2\kappa}A_{j}^{\rm in},
\\
\dot{b}_{j}=& -(i\omega _{m}+\gamma )b_{j}-it_{1}B_{j}-it_{2}B_{j-1}+iG_{j,1}a_{j}^{\dag }+iG_{j,1}^{*}a_{j}+\sqrt {2\gamma}b_{j}^{\rm in},
\\
\dot{B}_{j}=& -(i\omega _{m}+\gamma )B_{j}-it_{1}b_{j}-it_{2}b_{j+1}+iG_{j,2}A_{j}^{\dag }+iG_{j,2}^{*}A_{j}+\sqrt {2\gamma}B_{j}^{\rm in}.
\end{align}
\end{subequations}%
In Eqs.~(\ref{classical}) and (\ref{fluctuations}), we introduce the normalized detunings
\begin{equation}
 \Delta'_{a}\equiv\Delta_a -g_{1}(\beta_{j}^{*}+\beta _{j}),\qquad \Delta'_{A}\equiv\Delta_A -g_{2}(\eta_{j}^{*}+\eta _{j}),
\end{equation}
as well as the effective optomechanical coupling strengthes
\begin{equation}
G_{j,1}\equiv g_{1}\alpha _{j}, \qquad  G_{j,2}\equiv g_{2}\zeta _{j}
\end{equation}
at the $j$th site. In Eqs.~(\ref{fluctuations}), meanwhile, we neglected the nonlinear terms $ig_{1}a_{j}^{\dag
}a_{j}$, $ig_{2}A_{j}^{\dag}A_{j}$, $ig_{1}(b_{j}^{\dag }+b_{j})a_{j}$, and $ig_{2}(B_{j}^{\dag }+B_{j})A_{j}$.
Based on Eqs.~(\ref{fluctuations}), we can infer a linearized Hamiltonian
\begin{eqnarray}
H_{\text{lin}} &=& \sum_{j}(\Delta'_{a} a_{j}^{\dag }a_{j}+\Delta'_{A} A_{j}^{\dag
}A_{j})+\omega _{m}\sum_{j}(b_{j}^{\dag }b_{j}+B_{j}^{\dag }B_{j})  \notag \\
&&-\sum_{j}(G_{j,1}a_{j}^{\dag }+G_{j,1}^{*}a_{j})(b_{j}^{\dag }+b_{j})
-\sum_{j}(G_{j,2}A_{j}^{\dag }+G_{j,2}^{*}A_{j})(B_{j}^{\dag }+B_{j})  \notag \\
&&+J_{1}\sum_{j}(A_{j}^{\dag }a_{j}+a_{j}^{\dag
}A_{j})+J_{2}\sum_{j}(A_{j}^{\dag }a_{j+1}+a_{j+1}^{\dag }A_{j}) \notag\\
&&+t_{1}\sum_{j}(B_{j}^{\dag }b_{j}+b_{j}^{\dag
}B_{j})+t_{2}\sum_{j}(B_{j}^{\dag }b_{j+1}+b_{j+1}^{\dag }B_{j})
\end{eqnarray}
to govern the linearized dynamics of the system.

In this work, we consider the parameter conditions
\begin{equation}
\Delta_a \sim \Delta_A \sim \omega_m,\qquad |g_{1}(\beta_{j}^{*}+\beta _{j})|\ll\Delta_a,\qquad |g_{2}(\eta_{j}^{*}+\eta _{j})|\ll\Delta_A,
\end{equation}
then we have $\Delta'_a\approx\Delta_a$ and $\Delta'_A\approx\Delta_A$.
Furthermore, we consider the conditions
\begin{equation}
 \Delta_a \sim \Delta_A \sim \omega_m \gg \{J_1,J_2,t_1,t_2,G_{j,1},G_{j,2}\}.
\end{equation}
Under these conditions, the linearized Hamiltonian $H_{\rm lin}$ can be simplified by making the rotating-wave approximation (RWA) as
\begin{eqnarray}\label{Heff2}
H_{\text{lin}} &=&\sum_{j}(\Delta_a a_{j}^{\dag }a_{j}+ \Delta_A A_{j}^{\dag
}A_{j})+\omega _{m}\sum_{j}(b_{j}^{\dag }b_{j}+B_{j}^{\dag }B_{j})  \notag \\
&&-\sum_{j}G_{j,1}(a_{j}^{\dag }b_{j}+a_{j}b_{j}^{\dag })
-\sum_{j}G_{j,2}(A_{j}^{\dag }B_{j}+A_{j}B_{j}^{\dag })  \notag \\
&&+J_{1}\sum_{j}(A_{j}^{\dag }a_{j}+a_{j}^{\dag
}A_{j})+J_{2}\sum_{j}(A_{j}^{\dag }a_{j+1}+a_{j+1}^{\dag }A_{j}) \notag \\
&&+t_{1}\sum_{j}(B_{j}^{\dag }b_{j}+b_{j}^{\dag
}B_{j})+t_{2}\sum_{j}(B_{j}^{\dag }b_{j+1}+b_{j+1}^{\dag }B_{j}).
\end{eqnarray}
Here, $G_{j,1}$ and $G_{j,2}$ are considered as real numbers, which can be realized by choosing proper phases of the optical pumping fields. Below, we consider that all the linearized optomechanical couplings take the same coupling strength, i.e., $G_{j,1}=G_{j,2}=G$. Then the linearized Hamiltonian (\ref{Heff2}) becomes
\begin{eqnarray}
H_{\text{lin}} &=& \sum_{j}(\Delta_a a_{j}^{\dag }a_{j}+\Delta_A A_{j}^{\dag
}A_{j})+\omega _{m}\sum_{j}(b_{j}^{\dag }b_{j}+B_{j}^{\dag }B_{j})  \notag \\
&&-G\sum_{j}(a_{j}^{\dag }b_{j}+a_{j}b_{j}^{\dag }+A_{j}^{\dag }B_{j}+A_{j}B_{j}^{\dag }) \notag \\
&&+J_{1}\sum_{j}(A_{j}^{\dag }a_{j}+a_{j}^{\dag
}A_{j})+J_{2}\sum_{j}(A_{j}^{\dag }a_{j+1}+a_{j+1}^{\dag }A_{j}) \notag\\
&&+t_{1}\sum_{j}(B_{j}^{\dag }b_{j}+b_{j}^{\dag
}B_{j})+t_{2}\sum_{j}(B_{j}^{\dag }b_{j+1}+b_{j+1}^{\dag }B_{j}).
\end{eqnarray}
In a rotating frame defined by the unitary transformation operator $U_2=\exp[-i(\omega_{m}t)\sum_{j}(a_{j}^{\dag }a_{j}+A_{j}^{\dag }A_{j}+b_{j}^{\dag }b_{j}+B_{j}^{\dag }B_{j})]$, the Hamiltonian of the system becomes%
\begin{eqnarray}\label{Hsys}
H_{\text{sys}} &=&\delta\sum_{j}( a_{j}^{\dag }a_{j}- A_{j}^{\dag
}A_{j})  \notag \\
&&-G\sum_{j}(a_{j}^{\dag }b_{j}+a_{j}b_{j}^{\dag }+A_{j}^{\dag }B_{j}+A_{j}B_{j}^{\dag }) \notag \\
&&+J_{1}\sum_{j}(A_{j}^{\dag }a_{j}+a_{j}^{\dag
}A_{j})+J_{2}\sum_{j}(A_{j}^{\dag }a_{j+1}+a_{j+1}^{\dag }A_{j}) \notag\\
&&+t_{1}\sum_{j}(B_{j}^{\dag }b_{j}+b_{j}^{\dag
}B_{j})+t_{2}\sum_{j}(B_{j}^{\dag }b_{j+1}+b_{j+1}^{\dag }B_{j}),
\end{eqnarray}
where we introduce $\delta=\Delta_a-\omega_{m}=\omega_{m}-\Delta_A$ and assume $|\delta|\ll \omega_{m}$ hereafter.
This linearized Hamiltonian $H_{\text{sys}}$ in Eq.~(\ref{Hsys}) is the starting point of our studies in the main text.

Next, we discuss how to satisfy the parameter condition $G_{j,1}=G_{j,2}=G$, i.e., $g_{1}\alpha _{j}=g_{2}\zeta _{j}=G$.
Under the approximations $\Delta'_a\approx\Delta_a$ and $\Delta'_A\approx\Delta_A$, the equations for $\alpha _{j}$ and $\zeta _{j}$ are reduced to
\begin{subequations}
\begin{align}
-(i\Delta_a +\kappa )\alpha _{j}-iJ_{1}\zeta _{j}-iJ_{2}\zeta
_{j-1}-i\varepsilon _{j}=& 0, \\
-(i\Delta_A +\kappa )\zeta _{j}-iJ_{1}\alpha _{j}-iJ_{2}\alpha
_{j+1}-i\epsilon _{j}=& 0.
\end{align}%
\end{subequations}%
Below, we will consider the optomechanical ladder with two different kinds of boundaries: (i) periodic-boundary condition, and (ii) open-boundary condition.

Under the periodic-boundary condition with $N$ unit cells ($\alpha _{j+N}=\alpha _{j}$ and $\zeta _{j+N}=\zeta _{j}$), the constraints $g_{1}\alpha _{j}=g_{2}\zeta _{j}=G$ can be realized with $\alpha _{j}=\alpha$, $\zeta _{j}=\zeta$, and $\zeta=(g_1/g_2)\alpha$ for all the sites;
then the equations for $\alpha$ and $\zeta$ become
\begin{subequations}
\label{alpha1}
\begin{align}
-(i\Delta_a +\kappa )\alpha -iJ_{1}\zeta -iJ_{2}\zeta -i\varepsilon _{j}=& 0, \\
-(i\Delta_A +\kappa )\zeta -iJ_{1}\alpha -iJ_{2}\alpha-i\epsilon _{j}=& 0.
\end{align}%
\end{subequations}%
Therefore, the driving strengths of the pumping fields should satisfy the relations
\begin{subequations}
\label{epsilon1}
\begin{align}
\varepsilon _{j}=&\left[i\kappa-\Delta_a -(J_{1}+J_{2})\frac{g_1}{g_2}\right]\alpha, \\
\epsilon _{j}=&\left[\frac{g_1}{g_2}(i\kappa-\Delta_A ) -(J_{1}+J_{2})\right]\alpha,
\end{align}%
\end{subequations}%
for all the sites $j=0,1,\cdots,N-1$.

We can also derive the expressions of these driving strengths under the open-boundary condition with $N$ unit cells. For realizing $g_{1}\alpha _{j}=g_{2}\zeta _{j}=G$, concretely, we also need $\alpha _{j}=\alpha$, $\zeta _{j}=\zeta$, and $\zeta=(g_1/g_2)\alpha$ for all the sites; and the driving strengths $\varepsilon _{j}$ and $\epsilon _{j}$ of the lasers also satisfy the relations in Eqs.~(\ref{epsilon1}), except the driving strengths $\varepsilon _{0}$ and $\epsilon _{N-1}$ at the two open boundaries.
The equations for $\alpha$ and $\zeta$ at the two boundaries are given by
\begin{subequations}
\begin{align}
-(i\Delta_a +\kappa )\alpha -iJ_{1}\zeta  -i\varepsilon _{0}=& 0, \\
-(i\Delta_A +\kappa )\zeta -iJ_{1}\alpha  -i\epsilon _{N-1}=& 0.
\end{align}%
\end{subequations}
In this case, the driving strengths $\varepsilon _{0}$ and $\epsilon _{N-1}$ of the lasers at the two ending points are given by
\begin{subequations}
\begin{align}
\varepsilon _{0}=&\left(i\kappa-\Delta_a -J_{1}\frac{g_1}{g_2}\right)\alpha, \label{BC15a}\\
\epsilon _{N-1}=&\left[\frac{g_1}{g_2}(i\kappa-\Delta_A ) -J_{1}\right]\alpha.\label{BC15b}
\end{align}%
\end{subequations}%
Equations~(\ref{BC15a}) and (\ref{BC15b}) indicate that the driving amplitudes at the two ending points are different from those in Eq.~(\ref{epsilon1}) for the inner cavities.

\section{calculation of the boundary curves of the topological phases}

In this section, we calculate the energy bands for Hamiltonian (\ref{Hsys}) in the momentum space by introducing the discrete Fourier transformation and considering the periodic boundary conditions. We also derive the boundary curves of the topological phases based on the fact that topological phase transitions are always accompanied by the closing and reopening of the energy band gaps.

\subsection{Energy bands in the momentum space}

To calculate the energy bands of the system, we transform the Hamiltonian $H_{\text{sys}}$ in Eq.~(\ref{Hsys}) from the real space to the momentum space by introducing the bosonic operators $O_{k}$ in the momentum space as
\begin{equation}
O_{k}=\frac{1}{\sqrt{N}}\sum_{j}e^{ijkd_{0}}O_{j}\text{, \ \ \ \ \ }O=a,b,A,\text{ and }B,
\end{equation}%
where $k$ is the wave number and $d_{0}$ is the lattice constant (hereafter we set $d_{0}=1$ for simplicity). The Hamiltonian (\ref{Hsys}) can be transformed into the momentum space as
\begin{eqnarray}
H_{\text{sys}}
&=&\sum_{k}\left(
\begin{array}{llll}
a_{k}^{\dag } & A_{k}^{\dag } & b_{k}^{\dag } & B_{k}^{\dag }%
\end{array}%
\right) \left(
\begin{array}{cccc}
\delta  & J_{1}+J_{2}e^{ik} & -G & 0 \\
J_{1}+J_{2}e^{-ik} & -\delta  & 0 & -G \\
-G & 0 & 0  & t_{1}+t_{2}e^{ik} \\
0 & -G & t_{1}+t_{2}e^{-ik} & 0
\end{array}%
\right) \left(
\begin{array}{l}
a_{k} \\
A_{k} \\
b_{k} \\
B_{k}%
\end{array}%
\right) \notag\\
&=& \sum_{k}(V_{k})^{\dag }H_{k}V_{k},
\end{eqnarray}%
where we introduce $(V_{k})^{\dag }=(a_{k}^{\dag },A_{k}^{\dag },b_{k}^{\dag },B_{k}^{\dag })$ and
\begin{equation}
\label{Hk}
H_{k}=\left(
\begin{array}{cccc}
\delta  & J_{1}+J_{2}e^{ik} & -G & 0 \\
J_{1}+J_{2}e^{-ik} & -\delta  & 0 & -G \\
-G & 0 & 0 & t_{1}+t_{2}e^{ik} \\
0 & -G & t_{1}+t_{2}e^{-ik} & 0%
\end{array}%
\right).
\end{equation}%
The energy bands of the system can be obtained by diagonalizing $H_{k}$ in Eq.~(\ref{Hk}).

Under the resonant condition $\delta=0$, the energy bands of $H_{k}$ are given by%
\begin{subequations}
\begin{align}
E_{1}&=-\sqrt{G^{2}+\frac{1}{2}( \left\vert \rho _{2}\right\vert
^{2}+\left\vert \rho _{1}\right\vert ^{2}) +\sqrt{[G^{2}+\frac{1}{2}( \left\vert \rho _{2}\right\vert
^{2}+\left\vert \rho _{1}\right\vert ^{2})]^2-( G^{2}-\rho _{1}\rho _{2}^{\ast
}) \left( G^{2}-\rho _{1}^{\ast }\rho _{2}\right) }},\label{Ek1}\\
E_{2}&=-\sqrt{G^{2}+\frac{1}{2}( \left\vert \rho _{2}\right\vert
^{2}+\left\vert \rho _{1}\right\vert ^{2}) -\sqrt{[G^{2}+\frac{1}{2}( \left\vert \rho _{2}\right\vert
^{2}+\left\vert \rho _{1}\right\vert ^{2})]^2-\left( G^{2}-\rho _{1}\rho _{2}^{\ast
}\right) \left( G^{2}-\rho _{1}^{\ast }\rho _{2}\right) }},\label{Ek2}\\
E_{3}&=\sqrt{G^{2}+\frac{1}{2}( \left\vert \rho _{2}\right\vert
^{2}+\left\vert \rho _{1}\right\vert ^{2}) -\sqrt{[G^{2}+\frac{1}{2}( \left\vert \rho _{2}\right\vert
^{2}+\left\vert \rho _{1}\right\vert ^{2})]^2-\left( G^{2}-\rho _{1}\rho _{2}^{\ast
}\right) \left( G^{2}-\rho _{1}^{\ast }\rho _{2}\right) }},\label{Ek3}\\
E_{4}&=\sqrt{G^{2}+\frac{1}{2}( \left\vert \rho _{2}\right\vert
^{2}+\left\vert \rho _{1}\right\vert ^{2}) +\sqrt{[G^{2}+\frac{1}{2}( \left\vert \rho _{2}\right\vert
^{2}+\left\vert \rho _{1}\right\vert ^{2})]^2-\left( G^{2}-\rho _{1}\rho _{2}^{\ast
}\right) \left( G^{2}-\rho _{1}^{\ast }\rho _{2}\right) }}\label{Ek4}
\end{align}
\end{subequations}
with $\rho _{1}\equiv J_{1}+J_{2}e^{ik}$ and $\rho _{2}\equiv t_{1}+t_{2}e^{ik}$.

\subsection{Topological phase boundary curves}

\begin{figure}[tbp]
\includegraphics[bb=41 498 525 655, width=18cm, clip]{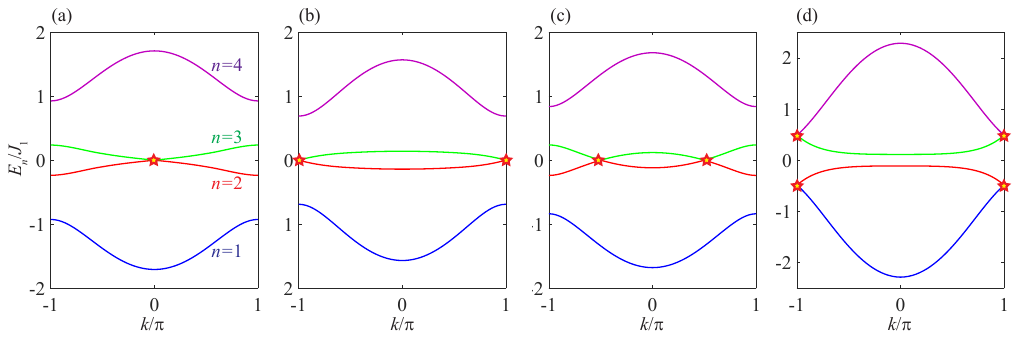}
\caption{(Color online) Energy bands of the optomechanical ladder for different optomechanical strengths: (a) $G=G_{c,1}$ and (b) $G=G_{c,2}$, when $J_{2}/J_{1}=0.5$, $t_{1}/J_{1}=0.2$, and $t_{2}/J_{1}=0.01$. (c) Energy bands of the optomechanical ladder when $G/J_{1}=0.5$, $t_{1}/J_{1}=0.2$, $t_{2}/J_{1}=0.1$, and $J_{2}/J_{1}=0.5$. (d) Energy bands of the optomechanical ladder when $t_{1}/J_{1}=0.2$, $t_{2}/J_{1}=0.01$, $J_{1}-J_{2}=t_{2}-t_{1}$, and $G/J_{1}=0.45$. There are four different energy spectra with closed energy bands: (a) $E_{2}=E_{3}=0$ at $k=0$; (b) $E_{2}=E_{3}=0$ at $k=\pm \pi$; (c) $E_{2}=E_{3}=0$ at $k=\pm \cos^{-1}[(G^2-J_1 t_1-J_2 t_2)/(J_2 t_1+J_1 t_2)]$; (d) $E_{1}=E_{2}$ and $E_{3}=E_{4}$ at $k=\pm \pi$.}
\label{FigS1}
\end{figure}

Based on Eqs.~(\ref{Ek1})-(\ref{Ek4}), we can obtain four energy bands in the energy spectra of the model, as shown in Figs.~\ref{FigS1}(a)-\ref{FigS1}(d). Typically, there are energy gaps between different bands, and the topological phase transition takes place when they close and reopen [Figs.~\ref{FigS1}(a)-\ref{FigS1}(d)]. Below, we will discuss the phase boundaries corresponding to these four cases.

First, the band gap closes for $E_{2}=E_{3}=0$ under the condition
\begin{equation}\label{TopoBan}
G^{2}-\rho _{1}\rho _{2}^{\ast }=0,
\end{equation}%
then we can obtain the equation for the boundaries as
\begin{equation}\label{Bd}
G=\sqrt{J_{1}t_{1}+J_{2}t_{2}+t_{1}J_{2}e^{ik}+J_{1}t_{2}e^{-ik}}.
\end{equation}%
In order to make sure that $G$ is real and positive, the band gap closes under different conditions for different $k$:

 \noindent (i) The band gap closes at $k=0$ [Fig.~\ref{FigS1}(a)], and the boundary in Eq.~(\ref{Bd}) becomes
\begin{equation}
G_{c,1}=\sqrt{\left( J_{1}+J_{2}\right) \left( t_{1}+t_{2}\right) }.
\end{equation}%
(ii) The band gap closes at $k=\pm \pi$ [Fig.~\ref{FigS1}(b)], and the corresponding boundary becomes
\begin{equation}
G_{c,2}=\sqrt{\left( J_{1}-J_{2}\right) \left( t_{1}-t_{2}\right) }.
\end{equation}%
(iii) The band gap closes at any other value in the regime $0<|k|<\pi$ [Fig.~\ref{FigS1}(c)], then the boundary is given by
\begin{equation} \label{Gc3}
G_{c,3}=\sqrt{J_{1}t_{1}+J_{2}t_{2}+\left( J_{2}t_{1}+J_{1}t_{2}\right) \cos k},
\end{equation}
with an additional precondition
\begin{equation}
t_{1}J_{2}=J_{1}t_{2}.
\end{equation}
It is worth emphasizing that the boundary of $G_{c,3}$ cannot be identified by the Berry phases, because the Berry phases do not change when the parameters cross this boundary, as shown in Figs.~1(c) and 1(e) in the main text.
The boundary of $G_{c,3}$ can be identified when two optomechanical ladders in different phases are attached, which is shown in the following subsection.

Second, the band gaps also close when $E_{1}=E_{2}$ and $E_{3}=E_{4}$ [Fig.~\ref{FigS1}(d)], which is satisfied under the condition
\begin{equation}\label{Jc2}
\left[G^{2}+\frac{1}{2}( \left\vert \rho _{2}\right\vert
^{2}+\left\vert \rho _{1}\right\vert ^{2})\right]^2-( G^{2}-\rho _{1}\rho
_{2}^{\ast }) ( G^{2}-\rho _{1}^{\ast }\rho _{2}) =0.
\end{equation}%
This condition can be simplified as
\begin{equation}
\rho _{1}+\rho _{2}=0,
\end{equation}%
which is satisfied when
\begin{equation}
e^{ik}=-\frac{J_{1}+t_{1}}{J_{2}+t_{2}}.
\end{equation}%
Since we assume that $J_{1}$, $J_{2}$, $t_{1}$, and $t_{2}$ are positive-real numbers, the boundary curve takes the following form%
\begin{equation} \label{J2c}
J_{2,c}=J_{1}+t_{1}-t_{2}
\end{equation}%
for $k= \pm \pi $.
All these equations determining the phase boundaries have been given in the main text.

\subsection{Generic symmetry operator and topological number}

The topological properties of the system governed by the Hamiltonian $H_k$ can be characterized by the Berry phase~\cite{Janos2016Springer}, and the \emph{topological phase diagrams of the optomechanical polaritons} based on the Berry phases are shown in Figs.~1(b) and 1(c) in the main text. In addition, the topological phases of one-dimensional lattices can also be characterized by the topological number associated with some generic symmetries. Following Refs.~\cite{Wakatsuki14PRB,Padavic2018PRB}, we will introduce the generic symmetry operator $S$ of the optomechanical ladder and the associated topological number $N_s$.

The generic symmetry operator for the optomechanical ladder is defined by%
\begin{equation}
S=\left(
\begin{array}{cccc}
1 & 0 & 0 & 0 \\
0 & -1 & 0 & 0 \\
0 & 0 & -1 & 0 \\
0 & 0 & 0 & 1%
\end{array}%
\right) .
\end{equation}%
It can be checked that
\begin{equation}
  S^{-1}H_k S=-H_k,\qquad S^{2}=\mathbb{I},
\end{equation}
where $\mathbb{I}$ is identity matrix.
Therefore, the topological class of the optomechanical ladder is the chiral orthogonal (BDI) class and its topological phases can be characterized by the $\mathbb{Z}$ index~\cite{Schnyder08PRB}.
The topological number associated with the generic symmetry operator $S$ is
defined by%
\begin{equation}\label{NsDefi}
N_{s}=\mathrm{Tr}\left[ \int_{-\pi }^{+\pi }\frac{dk}{4\pi i}Sg_s^{-1}\left(
k\right) \partial _{k}g_s\left( k\right) \right],
\end{equation}%
where $g_s\left( k\right) =-H_k^{-1} $ is the Green's function at zero energy.

To make the calculation of the topological number much easier, we introduce a unitary operator
\begin{equation}\label{Eq_Us}
U_{s}=\left(
\begin{array}{cccc}
1 & 0 & 0 & 0 \\
0 & 0 & 0 & 1 \\
0 & 0 & 1 & 0 \\
0 & 1 & 0 & 0%
\end{array}%
\right) ,
\end{equation}%
which yields
\begin{equation}
U_{s}SU_{s}^{-1}=\sigma _{z}\otimes \mathbb{I}
\end{equation}
with the Pauli matrix $\sigma _{z}$ and $2\times 2$ identity matrix $\mathbb{I}$.
Based on Eq.~(\ref{Eq_Us}), it can be shown that
\begin{equation}\label{EqHkV}
U_{s}H_k U_{s}^{-1}=\left(
\begin{array}{cc}
0 & V \\
V^{\dag } & 0%
\end{array}%
\right) ,
\end{equation}%
with%
\begin{equation}
V=\left(
\begin{array}{cc}
-G & \rho _{1}  \\
\rho _{2}^{\ast } & -G%
\end{array}%
\right).
\end{equation}
The topological number for the Hamiltonian with the form of Eq.~(\ref{EqHkV}) is given by%
\begin{eqnarray}
N_{s} &=&-\mathrm{Tr}\left[ \int_{-\pi }^{+\pi }\frac{dk}{2\pi i}%
V^{-1}\partial _{k}V\right]   \nonumber \\
&=&-\int_{-\pi }^{+\pi }\frac{dk}{2\pi i}\partial _{k}\{\ln [\mathrm{Det}(V)]\}
\nonumber \\
&=&-\int_{-\pi }^{+\pi }\frac{dk}{2\pi i}\partial _{k}\{\ln [Z(k)]\},
\end{eqnarray}%
where $Z(k)= G^{2}-\rho_1 \rho_2 ^{\ast }$. It means that $N_s$ is the winding number of $Z(k)$ in the complex plane.
The gap closes with $Z(k)=0$, which is consistent with the phase boundaries given by Eq.~(\ref{TopoBan}).
We should emphasize that the gaps also close with $J_{2,c}$ at nonzero energy [$E_n \neq 0$, see Fig.~\ref{FigS1}(d)], which cannot be described by the topological number $N_s$ defined based on the Green's function at zero energy [Eq.~(\ref{NsDefi})].

\subsection{Verification of the phase boundary curves $G_{3,c}$ and $J_{c,2}$}

\begin{figure}[tbp]
\includegraphics[width=0.98 \textwidth]{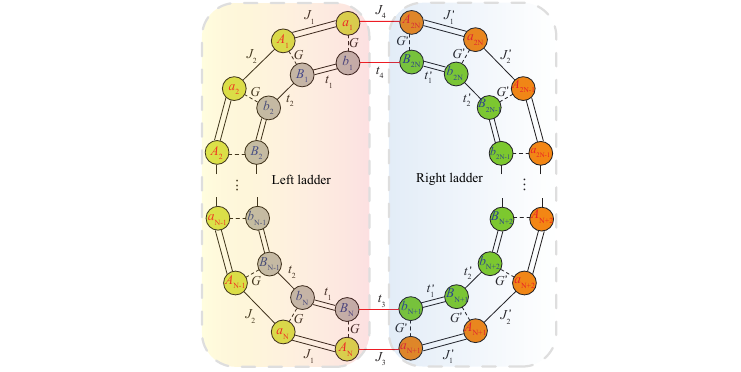}
\caption{(Color online) Schematic of the optomechanical-ladder ring consisting of two optomechanical ladders. The coupling parameters within the left (right) ladder are $J_{1}$, $t_{1}$, $J_{2}$, $t_{2}$, and  $G$ ($J_{1}^{\prime}$, $t_{1}^{\prime}$, $J_{2}^{\prime}$, $t_{2}^{\prime}$, and  $G^{\prime}$). The coupling strengths between the left and right optomechanical ladders are $J_{3}$, $t_{3}$, $J_{4}$, and $t_{4}$.}
\label{FigS2}
\end{figure}

\begin{figure}[tbp]
\includegraphics[width=0.95 \textwidth]{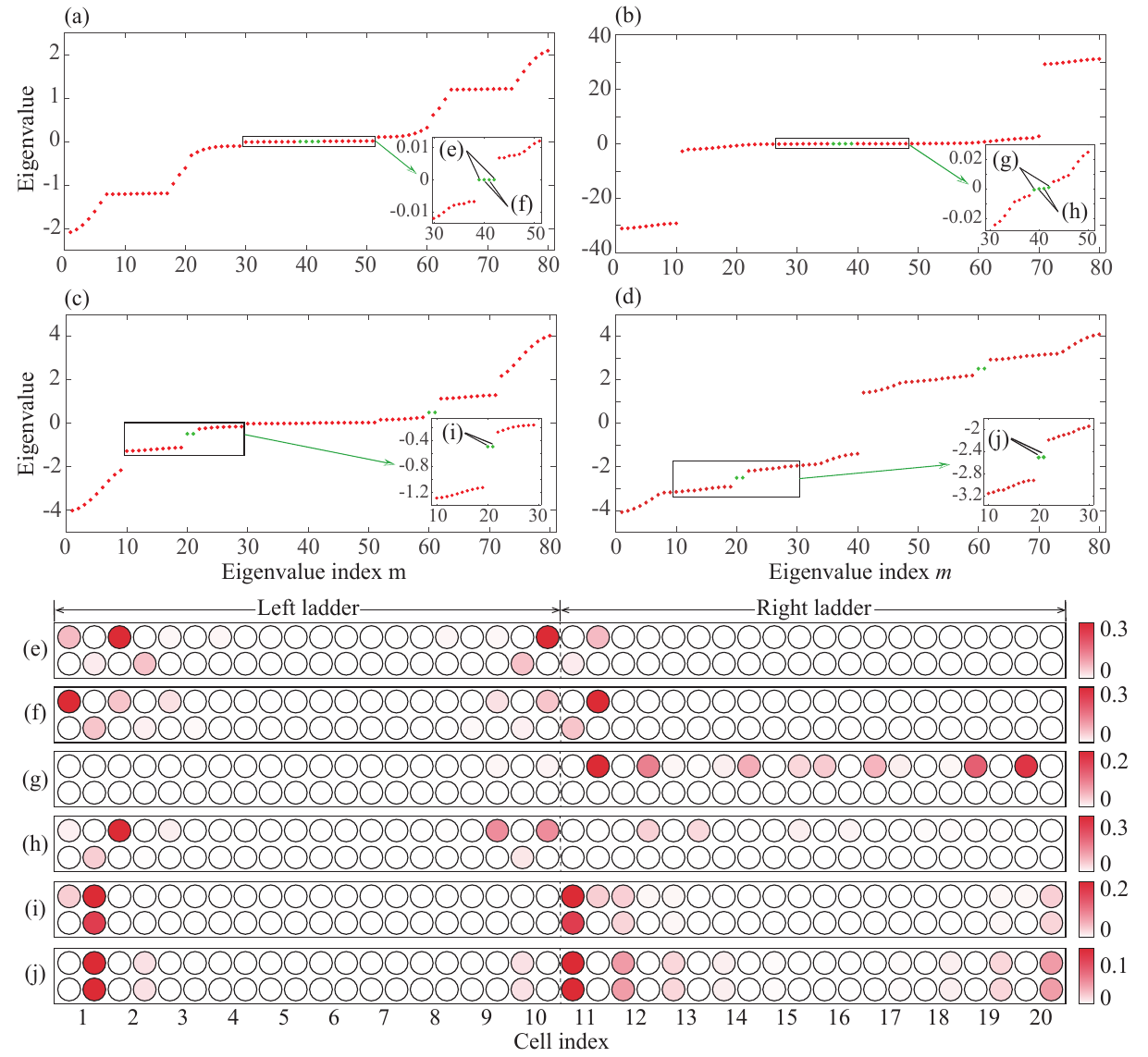}
\caption{(Color online) The eigenvalues of the optomechanical-ladder ring ($N = 10$ cells in one ladder) for (a) $J_{2}/J_{1}=0.01$, $J_{1}^{\prime}/J_{1}=1$, $J_{2}^{\prime}/J_{1}=1$,  $t_{1}/J_{1}=t_{1}^{\prime}/J_{1}=0.2$, $t_{2}/J_{1}=t_{2}^{\prime}/J_{1}=0.01$, and $G/J_{1}=G^{\prime}/J_{1}=0.45$; (b) $J_{2}/J_{1}=1$, $J_{1}^{\prime}/J_{1}=1$, $J_{2}^{\prime}/J_{1}=30$,  $t_{1}/J_{1}=t_{1}^{\prime}/J_{1}=0.01$, $t_{2}/J_{1}=t_{2}^{\prime}/J_{1}=0.2$, $G/J_{1}=0.5$, and $G^{\prime}/J_{1}=2.5$; (c) $J_{2}/J_{1}=0.01$, $J_{1}^{\prime}/J_{1}=1$, $J_{2}^{\prime}/J_{1}=1$,  $t_{1}/J_{1}=t_{1}^{\prime}/J_{1}=0.2$, $t_{2}/J_{1}=t_{2}^{\prime}/J_{1}=0.01$, and $G/J_{1}=G^{\prime}/J_{1}=0.45$; (d) $J_{2}/J_{1}=0.1$, $J_{1}^{\prime}/J_{1}=1$, $J_{2}^{\prime}/J_{1}=1.5$,  $t_{1}/J_{1}=t_{1}^{\prime}/J_{1}=0.01$, $t_{2}/J_{1}=t_{2}^{\prime}/J_{1}=0.2$, and $G/J_{1}=G^{\prime}/J_{1}=2.5$. (e-j) The field distribution of the edge states in the eigenvalue spectra (green dots). }
\label{FigS3}
\end{figure}

In the above subsections, we have obtained the four topological phase boundary curves ($G_{c,1}$, $G_{c,2}$, $G_{c,3}$, and $J_{2,c}$) based on the fact that topological phase transition takes place when two of the energy band gaps are closed and reopened. However, the boundary $G_{c,3}$ cannot be described by the Berry phases [see Figs.1(b) and 1(c) in the main text], and the boundary $J_{2,c}$ cannot be described by the topological number $N_s$.
To confirm the topological phase boundary $G_{3,c}$ [Eq.~(\ref{Gc3})] and $J_{c,2}$ [Eq.~(\ref{J2c})] in the optomechanical ladder, in this subsection, we consider an optomechanical-ladder ring (Fig.~\ref{FigS2}) consisting of two coupled optomachanical ladders with parameters belong to different regimes divided by the phase boundary $G_{3,c}$ or $J_{c,2}$.
The Hamiltonian of the optomechanical-ladder ring reads
\begin{eqnarray}
H_{\text{ring}} &=&\sum_{j=1}^{N-1}(J_{1}A_{j}^{\dag
}a_{j}+J_{2}a_{j+1}^{\dag }A_{j}+t_{1}B_{j}^{\dag }b_{j}+t_{2}B_{j}^{\dag
}b_{j+1}+\text{H.c.})-G \sum_{j=1}^{N}(a_{j}^{\dag }b_{j}+A_{j}^{\dag }B_{j}+\text{H.c.}%
) \notag \\
&&+\sum_{j=N+1}^{2N-1}(J_{1}^{\prime }A_{j}^{\dag }a_{j}+J_{2}^{\prime
}a_{j+1}^{\dag }A_{j}+t_{1}^{\prime }B_{j}^{\dag }b_{j}+t_{2}^{\prime
}B_{j}^{\dag }b_{j+1}+\text{H.c.})-G^{\prime }\sum_{j=N+1}^{2N}(a_{j}^{\dag }b_{j}+A_{j}^{\dag }B_{j}+\text{H.c.}%
) \notag \\
&&+(J_{3}A_{N}^{\dag }a_{N+1}+t_{3}B_{N}^{\dag }b_{N+1}+\text{H.c.})+(J_{4}A_{2N}^{\dag }a_{1}+t_{4}B_{2N}^{\dag }b_{1}+\text{H.c.}),
\end{eqnarray}%
where $J_{1}$, $J_{2}$, $t_{1}$, $t_{2}$, and $G$ ($J_{1}^{\prime }$, $J_{2}^{\prime }$, $t_{1}^{\prime }$, $t_{2}^{\prime }$, and $G^{\prime }$) are the coupling strengths for the left (right) optomechanical ladder. The parameters $J_{3}$, $J_{4}$, $t_{3}$, and $t_{4}$ are the coupling strengths between the two optomechanical ladders.
The topological phase transition boundaries $G_{c,3}$ and $J_{2,c}$ can be confirmed by analyzing the edge states and the corresponding field distributions of the optomechanical-ladder ring when the parameters cross the boundaries.
We note that when the optomechanical ladders belong to different topological phases, there are edge states in the eigenvalue spectra of the optomechanical-ladder ring. On the contrary, if there are no edge states in the eigenvalue spectra of the ring, it means that the two optomechanical ladders are in the same topological phase.
The eigenvalues and the corresponding field distributions of edge states for the optomechanical-ladder ring are shown in Fig.~\ref{FigS3}.

When both the left and right optomechanical ladders are in the topological phase II [see Fig. 1(b) in the main text] but the coupling parameters belong to the left and right sides of the boundary $G_{c,3}$ respectively, there are four degenerate edge states around the zero energy shown by green dots in Fig.~\ref{FigS3}(a).
The field distributions of the edge states are shown in Figs.~\ref{FigS3}(e) and \ref{FigS3}(f). The maxima of the fields in Figs.~\ref{FigS3}(e) and \ref{FigS3}(f) for the edge states are located around the connection region of the ring, which indicate that the phase transitions take place when the parameters cross the phase boundary $G_{c,3}$.
In addition, when both of the coupling parameters of the left and right ladders are in the topological phase III [see Fig. 1(c) in the main text] and the coupling parameters belong to both sides of the boundary $G_{c,3}$, there are also four degenerate edge states with zero energy shown in Fig.~\ref{FigS3}(b) and the field distributions of the edge states are shown in Figs.~\ref{FigS3}(g) and \ref{FigS3}(h). Different from the edge states shown in Figs.~\ref{FigS3}(e) and \ref{FigS3}(f), most of the edge fields are located at one of the optomechanical ladders and around the connection region of the ring, which also indicate that the phase transitions take place when the parameters cross the phase boundary $G_{c,3}$.

We also discuss the eigenvalues of the optomechanical-ladder ring with the coupling parameter $J_2<J_{2,c}$ in the left ladder (in the phase I or II) and $J^{\prime}_2>J_{2,c}$ in the right ladder (in the phase III or IV). When the left and right optomechanical ladders are in the phase II and phase III [divided by the boundary $J_{2,c}$ shown in Fig.1(b)] respectively, there are four edges states (two at $E\approx - G$ and two at $E\approx G$), as shown in Fig.~\ref{FigS3}(c). Similarly, when the left and right optomechanical ladders are located in the phase I and phase IV [divided by the boundary $J_{2,c}$ shown in Fig. 1(c)] respectively, there are also four edges states around $E\approx \pm G$ [see Fig.~\ref{FigS3}(d)]. The maxima of the fields are located at the connection region, as shown in Figs.~\ref{FigS3}(i)-\ref{FigS3}(l). This means that phase transitions also take place when the parameters cross the phase boundary $J_{2,c}$.

\section{detection of the edge states}

\begin{figure}[tbp]
\includegraphics[width=0.98 \textwidth]{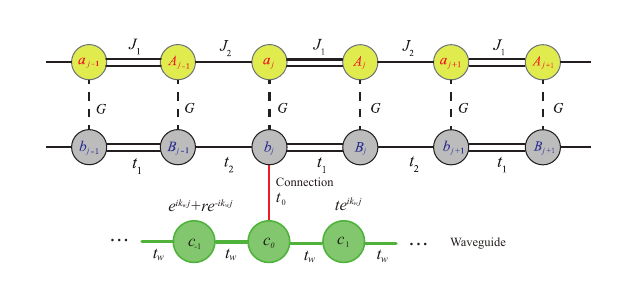}
\caption{(Color online) Schematic of the system connected with the waveguide. The waveguide is assumed as a chain of cavities shown by green circles with intracellular coupling strength $t_{w}$. The red line stands for the connection between the system and the waveguide with coupling strength $t_{0}$.}
\label{FigS4}
\end{figure}

In this section, we discuss how to \emph{detect the edge states in the optomechanical ladder} with an open boundary. Concretely, we introduce a one-dimensional (1D) phonon waveguide side-coupled to the optomechanical ladder, as shown in Fig.~\ref{FigS4}. The edge states in the optomechanical ladder can be detected by analyzing the \emph{reflection of a single phonon} transported in the 1D phonon waveguide~\cite{Zhou2008PRL,Zhou2008PRA,Zhou2009PRA,Liao2010PRA}.

We assume that the waveguide is coupled to the optomechanical ladder through the mechanical mode $c_{0}$. In a rotating frame defined by the unitary transformation operator $U_3=\exp[-i(\omega_{m}t)\sum_{j}(a_{j}^{\dag }a_{j}+A_{j}^{\dag }A_{j}+b_{j}^{\dag }b_{j}+B_{j}^{\dag }B_{j}+c_{j}^{\dag }c_{j})]$, the whole system including both the optomechanical ladder and the waveguide can be described by the total Hamiltonian
\begin{equation}
H_{\rm tot}=H_{\rm sys}+H_{\rm wg}+H_{\rm int},
\end{equation}
where $H_{\rm sys}$ [defined in Eq.~(\ref{Hsys})] is the Hamiltonian of the optomechanical ladder, $H_{\rm wg}$ is the waveguide Hamiltonian defined by
\begin{equation}\label{e29}
H_{\rm wg}=-t_{w}\sum_{j}( c_{j}^{\dag}c_{j+1}+ c_{j+1}^{\dag}c_{j}),
\end{equation}%
and the interaction Hamiltonian between the waveguide and the optomechanical ladder reads
\begin{equation}\label{eq30}
H_{\rm int}=t_{0}( c_{0}^{\dag }d+c_{0}d^{\dag }),
\end{equation}%
with the coupling strength $t_0$.
In Eq.~(\ref{e29}), $\omega _{m}$ is the resonance frequency of all the mechanical modes, $c_j$ is the annihilation operator of the mechanical mode at the $j$th site in the waveguide, and $t_{w}$ is the coupling strength between two nearest-neighboring mechanical modes. In Eq.~(\ref{eq30}), $d$ is the annihilation operator for the mechanical mode coupled to the waveguide, i.e., depending on the coupling position, $d=\{b_{0}, B_{0}, b_{1}, B_{1},..., b_{N-1}, B_{N-1}\}$.

The stationary state of single-photon scattering in the system can be written as
\begin{equation}
\left\vert \Omega\right\rangle =\sum_{j}\left[ u_{a}\left( j\right)
a_{j}^{\dagger }+u_{A}\left( j\right) A_{j}^{\dagger }+u_{b}\left( j\right)
b_{j}^{\dagger }+u_{B}\left( j\right) B_{j}^{\dagger }+u_{c}\left( j\right)
c_{j}^{\dagger }\right] \left\vert \emptyset \right\rangle ,
\end{equation}%
where $\left\vert \emptyset\right\rangle $ is the vacuum state of the whole system, and $u_{c}\left( j\right) $ and $u_{O}\left( j\right)$ are the probability amplitudes corresponding to a single photon or phonon in the modes $c_{j}$ and $O_{j}$ ($O=a,A,b,B$), respectively.
The dispersion relation of the 1D (infinite site) waveguide is given by~\cite{Zhou2008PRL}
\begin{equation}
 \Omega _{k}=-2t_{w}\cos k_{w}, \qquad -\pi<k_{w}\leq \pi,
\end{equation}
where $\Omega_{k}$ is the energy of the input single photon and $k_{w}$ is the corresponding wave number.

\begin{figure}[tbp]
\includegraphics[width=0.98 \textwidth]{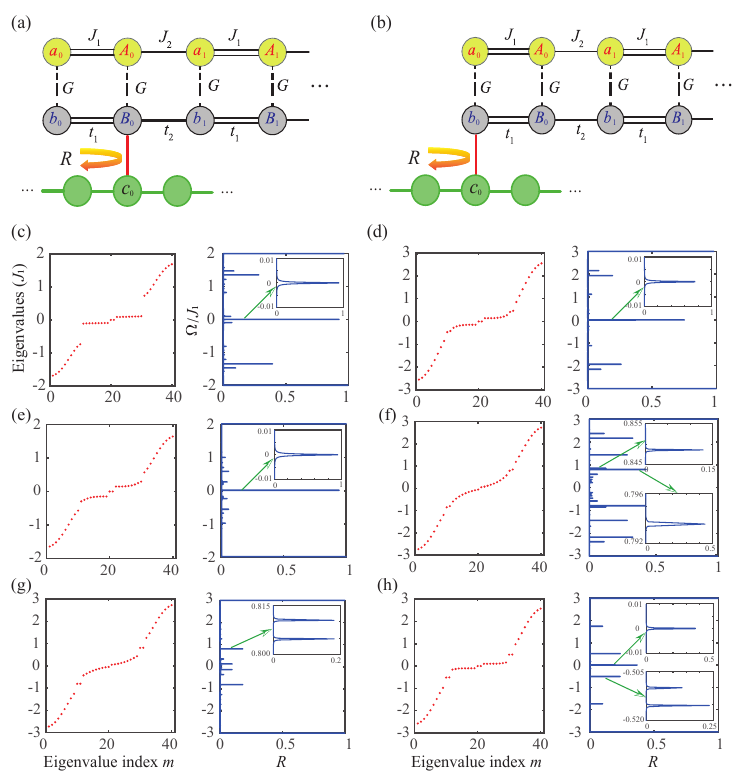}
\caption{(Color online) (a,b) Schematic of two different connections between the system and the waveguide: (a) the waveguide cavity $c_{0}$ is connected to $B_{0}$; (b) the waveguide cavity $c_{0}$ is connected to $b_{0}$. (c-h) The eigenvalues (red dots) and the corresponding reflection probability $R$ (blue solid curve) as functions of the energy, for an open-boundary condition with $N=10$ in various cases:
(c,d) the waveguide is connected to $B_{0}$ for (c) $J_{2}/J_{1}=0.6$, $t_{1}/J_{1}=0.2$, $t_{2}/J_{1}=0.01$, $G/J_{1}=0.4$, $t_{0}/J_{1}=0.025$, $t_{w}/J_{1}=1$ and (d) $J_{2}/J_{1}=1.5$, $t_{1}/J_{1}=0.2$, $t_{2}/J_{1}=0.01$, $G/J_{1}=0.4$, $t_{0}/J_{1}=0.025$, $t_{w}/J_{1}=1.5$;
(e-h) the waveguide is connected to $b_{0}$ for (e) $J_{2}/J_{1}=0.6$, $t_{1}/J_{1}=0.01$, $t_{2}/J_{1}=0.2$, $G/J_{1}=0.3$, $t_{0}/J_{1}=0.025$, $t_{w}/J_{1}=1$, (f) $J_{2}/J_{1}=1.5$, $t_{1}/J_{1}=0.2$, $t_{2}/J_{1}=0.01$, $G/J_{1}=0.8$, $t_{0}/J_{1}=0.025$, $t_{w}/J_{1}=1.5$, (g) $J_{2}/J_{1}=1.5$, $t_{1}/J_{1}=0.01$, $t_{2}/J_{1}=0.2$, $G/J_{1}=0.8$, $t_{0}/J_{1}=0.025$, $t_{w}/J_{1}=1.5$,
  and (h) $J_{2}/J_{1}=1.5$, $t_{1}/J_{1}=0.01$, $t_{2}/J_{1}=0.2$, $G/J_{1}=0.5$, $t_{0}/J_{1}=0.025$, $t_{w}/J_{1}=1.5$.}
\label{FigS5}
\end{figure}

According to the different distributions of the edge states as shown in Figs. 2(e)-2(j) in the main text, we consider two different connection situations between the optomechanical ladder and the waveguide, as shown in Figs.~\ref{FigS5}(a) and~\ref{FigS5}(b).

In order to detect the edge states with the maxima of the field in $B_{0}$ as shown in Figs. 2(e) and 2(i), we assume that the waveguide is coupled to the mechanical mode $B_{0}$ [Fig.~\ref{FigS5}(a)], i.e., $d=B_0$, and
the interaction Hamiltonian $H_{\rm int}$ reads
\begin{equation}
H_{\rm int}=t_{0}( c_{0}^{\dag }B_{0}+c_{0}B_{0}^{\dag }).
\end{equation}%
By substituting the stationary state $|\Omega\rangle$ and the Hamiltonian $H_{\rm tot}$ into the Schr\"{o}dinger equation $H_{\text{tot}}\left\vert \Omega\right\rangle =\Omega \left\vert \Omega\right\rangle $, we can obtain the coupled equations for the probability amplitudes as follows.
(i) When $0<j<N-1$ (in the bulk of the optomechanical ladder), the equations are given by
\begin{subequations}\label{A1}
\begin{align}
\Omega u_{a}(j) -J_{1}u_{A}(j) -J_{2}u_{A}(j-1) +Gu_{b}(j) =& 0,  \\
\Omega u_{b}(j) -t_{1}u_{B}(j) -t_{2}u_{B}(j-1) +Gu_{a}(j) =& 0, \\
\Omega u_{A}(j) -J_{1}u_{a}(j) -J_{2}u_{a}(j+1) +Gu_{B}(j) =& 0,  \\
\Omega u_{B}(j) -t_{1}u_{b}(j) -t_{2}u_{b}(j+1) +Gu_{A}(j) =& 0.
\end{align}
\end{subequations}
(ii) When $j=0$ and $j=N-1$ (at the boundaries of the optomechanical ladder), the equations take the form
\begin{subequations}\label{A2}
\begin{align}
\Omega u_{a}(0) -J_{1}u_{A}(0) +Gu_{b}(0) =& 0, \\
\Omega u_{b}(0) -t_{1}u_{B}(0) +Gu_{a}(0) =& 0, \\
\Omega u_{A}(0) -J_{1}u_{a}(0) -J_{2}u_{a}(1) +Gu_{B}(0) =& 0, \\
\Omega u_{B}(0) -t_{1}u_{b}(0) -t_{2}u_{b}(1) -t_{0}u_{c}(0) +Gu_{A}(0) =& 0,\\
\Omega u_{a}(N-1) -J_{1}u_{A}(N-1) -J_{2}u_{A}(N-2) +Gu_{b}(N-1) =& 0,\\
\Omega u_{b}(N-1) -t_{1}u_{B}(N-1) -t_{2}u_{B}(N-2) +Gu_{a}(N-1) =& 0, \\
\Omega u_{A}(N-1) -J_{1}u_{a}(N-1) +Gu_{B}(N-1) =& 0,\\
\Omega u_{B}(N-1) -t_{1}u_{b}(N-1) +Gu_{A}(N-1) =& 0, \\
\Omega u_{c}(0) +t_{w}u_{c}(1) +t_{w}u_{c}(-1) -t_{0}u_{B}(0) =& 0.
\end{align}
\end{subequations}
Moreover, for the waveguide, the probability amplitudes are determined by the following equations
\begin{equation}\label{wgap}
\Omega  u_{c}\left( j\right) +t_{w}u_{c}\left( j+1\right) +t_{w}u_{c}\left( j-1\right) =0
\end{equation}
for $j\neq 0$.

When a single phonon with energy $\Omega=-2t_{w}\cos k_{w}$ is injected from the left of the waveguide, a
general expression of the probability amplitudes in the waveguide is given by~\cite{Zhou2008PRL}%
\begin{equation}
u_{c}\left( j\right) =\left\{
\begin{array}{ll}
e^{ijk_{w}}+re^{-ijk_{w}}, & \qquad j<0, \\
te^{ijk_{w}},& \qquad j>0,%
\end{array}%
\right.
\end{equation}%
where $r$ and $t$ are, respectively, the reflection and transmission amplitudes, which satisfy both the probability conservation $\left\vert r\right\vert ^{2}+\left\vert t\right\vert ^{2}=1$ and the condition of continuity $t=1+r$. Finally, we obtain the coupled equations for the reflection amplitude $r$ and the probability amplitudes $u_{O}(j)$ in the optomechanical ladder, as
\begin{equation}
\label{matrix1}
\left(
\begin{array}{ccccccccccl}
\Omega  & G & -J_{1} & 0 & 0 & 0 & 0 & 0 & 0 & 0 & ... \\
G & \Omega  & 0 & -T_1 & 0 & 0
& 0 & 0 & 0 & 0 & ... \\
-J_{1} & 0 & \Omega  & T_G & -J_{2}
& 0 & 0 & 0 & 0 & 0 & ... \\
0 & -t_{1} & G & T_0 & 0 & -t_{2} & 0 & 0 & 0 & 0 & ... \\
0 & 0 & -J_{2} & 0 & \Omega  & G & -J_{1} & 0 & 0 & 0 & ... \\
0 & 0 & 0 & -t_{2} & G & \Omega  & 0 & -t_{1} & 0 & 0 & ... \\
0 & 0 & 0 & 0 & -J_{1} & 0 & \Omega  & G & -J_{2} & 0 & ... \\
0 & 0 & 0 & 0 & 0 & -t_{1} & G & \Omega  & 0 & -t_{2} & ... \\
\multicolumn{1}{c}{...} & \multicolumn{1}{c}{...} & \multicolumn{1}{c}{...}
& \multicolumn{1}{c}{...} & \multicolumn{1}{c}{...} & \multicolumn{1}{c}{...}
& \multicolumn{1}{c}{...} & \multicolumn{1}{c}{...} & \multicolumn{1}{c}{...}
& \multicolumn{1}{c}{...} & ...%
\end{array}%
\right) \left(
\begin{array}{l}
u_{a}\left( 0\right)  \\
u_{b}\left( 0\right)  \\
u_{A}\left( 0\right)  \\
r \\
u_{a}\left( 1\right)  \\
u_{b}\left( 1\right)  \\
u_{A}\left( 1\right)  \\
u_{B}\left( 1\right)  \\
...%
\end{array}%
\right) =\left(
\begin{array}{l}
0 \\
0 \\
0 \\
t_{0} \\
0 \\
0 \\
0 \\
0 \\
...%
\end{array}%
\right),
\end{equation}
where
\begin{equation}
T_1=2i(t_{w}t_{1}/t_{0})\sin (k_{w}), \quad T_G=2i(t_{w}/t_{0})G\sin (k_{w}), \quad T_0=-4i(t_{w}^{2}/t_{0})\cos (k_{w})\sin (k_{w}) -t_{0}.
\end{equation}
Then the reflection probability $R\equiv |r|^2$ can be obtained by solving Eq.~(\ref{matrix1}) numerically.

In order to detect the edge states with the maximal field at site $b_{0}$, as shown in Figs.~2(f)-2(h) in the main text, we assume that the waveguide is coupled to the mechanical mode $b_{0}$ [Fig.~\ref{FigS5}(b)], i.e., $d=b_0$, then
the interaction Hamiltonian $H_{\rm int}$ becomes
\begin{equation}
H_{\rm int}=t_{0}( c_{0}^{\dag }b_{0}+c_{0}b_{0}^{\dag }).
\end{equation}%
It is worth mentioning that the edge state in Fig.~2(j) can also be detected by coupling the waveguide to the mechanical mode $b_{0}$. This is because the field site $b_{0}$ is sufficiently strong (although not the strongest) to create the signal response.
By using a similar method, we can obtain the coupled equations for the probability amplitudes as follows:\\
(i) When $0<j<N-1$, we obtain
\begin{subequations}
\begin{align}
\Omega u_{a}(j) -J_{1}u_{A}(j) -J_{2}u_{A}(j-1) +Gu_{b}(j) =& 0, \\
\Omega u_{b}(j) -t_{1}u_{B}(j) -t_{2}u_{B}(j-1) +Gu_{a}(j) =& 0, \\
\Omega u_{A}(j) -J_{1}u_{a}(j) -J_{2}u_{a}(j+1) +Gu_{B}(j) =& 0, \\
\Omega u_{B}(j) -t_{1}u_{b}(j) -t_{2}u_{b}(j+1) +Gu_{A}(j) =& 0.
\end{align}
\end{subequations}
(ii) When $j=0$ and $j=N-1$, the equations are given by
\begin{subequations}
\begin{align}
\Omega u_{a}(0) -J_{1}u_{A}(0) +Gu_{b}(0) =& 0, \\
\Omega u_{b}(0) -t_{1}u_{B}(0) -t_{0}u_{c}(0) +Gu_{a}(0)=& 0,\\
\Omega u_{A}(0) -J_{1}u_{a}(0) -J_{2}u_{a}(1) +Gu_{B}(0) =& 0, \\
\Omega u_{B}(0) -t_{1}u_{b}(0) -t_{2}u_{b}(1) +Gu_{A}(0) =& 0,\\
\Omega u_{a}(N-1) -J_{1}u_{A}(N-1) -J_{2}u_{A}(N-2) +Gu_{b}(N-1) =& 0,\\
\Omega u_{b}(N-1)-t_{1}u_{B}(N-1)-t_{2}u_{B}(N-2)+Gu_{a}(N-1) =& 0,\\
\Omega u_{A}(N-1) -J_{1}u_{a}(N-1) +Gu_{B}(N-1) =& 0,\\
\Omega u_{B}(N-1) -t_{1}u_{b}(N-1) +Gu_{A}(N-1) =& 0,\\
\Omega u_{c}(0) +t_{w}u_{c}(1) +t_{w}u_{c}(-1) -t_{0}u_{b}(0) =& 0.
\end{align}
\end{subequations}
In addition, the coupled equations for the probability amplitudes in the waveguide are the same as Eq.~(\ref{wgap}).
Finally, the reflection amplitude $r$ and the probability amplitudes $u_{O}(j)$ in the optomechanical ladder satisfy the equation
\begin{equation}
\label{matrix2}
\left(
\begin{array}{ccccccccccl}
\Omega  & T_{G} & -J_{1} & 0 & 0 & 0
& 0 & 0 & 0 & 0 & ... \\
G & T_{0} & 0 & -t_{1} & 0 & 0 & 0 & 0 & 0 & 0 & ... \\
-J_{1} & 0 & \Omega  & G & -J_{2} & 0 & 0 & 0 & 0 & 0 & ... \\
0 & -T_{1} & G & \Omega  & 0 &
-t_{2} & 0 & 0 & 0 & 0 & ... \\
0 & 0 & -J_{2} & 0 & \Omega  & G & -J_{1} & 0 & 0 & 0 & ... \\
0 & 0 & 0 & -t_{2} & G & \Omega  & 0 & -t_{1} & 0 & 0 & ... \\
0 & 0 & 0 & 0 & -J_{1} & 0 & \Omega  & G & -J_{2} & 0 & ... \\
0 & 0 & 0 & 0 & 0 & -t_{1} & G & \Omega  & 0 & -t_{2} & ... \\
\multicolumn{1}{c}{...} & \multicolumn{1}{c}{...} & \multicolumn{1}{c}{...}
& \multicolumn{1}{c}{...} & \multicolumn{1}{c}{...} & \multicolumn{1}{c}{...}
& \multicolumn{1}{c}{...} & \multicolumn{1}{c}{...} & \multicolumn{1}{c}{...}
& \multicolumn{1}{c}{...} & ...%
\end{array}%
\right) \left(
\begin{array}{l}
u_{a}\left( 0\right)  \\
r \\
u_{A}\left( 0\right)  \\
u_{B}\left( 0\right)  \\
u_{a}\left( 1\right)  \\
u_{A}\left( 1\right)  \\
u_{b}\left( 1\right)  \\
u_{B}\left( 1\right)  \\
...%
\end{array}%
\right) =\left(
\begin{array}{l}
0 \\
t_{0} \\
0 \\
0 \\
0 \\
0 \\
0 \\
0 \\
...%
\end{array}%
\right).
\end{equation}
The reflection probability $R$ can be obtained by numerically solving Eq.~(\ref{matrix2}).

The eigenvalues and the corresponding reflection spectra are shown in Figs.~\ref{FigS5}(c)-\ref{FigS5}(h).
It can be seen from Figs.~\ref{FigS5}(c) and \ref{FigS5}(d) that, there are two degenerate-discrete states at zero energy, which correspond to the edge states with the field distribution shown in Figs.~2(e) and 2(i) in the main text.
There is a high peak in the corresponding reflection spectra at zero energy when the waveguide is coupled to mechanical mode $B_{0}$.
Instead, the two edge states at zero energy in Fig.~\ref{FigS5}(e) are located around the modes $b_{0}$ and $B_{N-1}$, so a high peak in the reflection spectra is detected when the waveguide is coupled to the mechanical mode $b_{0}$.

Similarly, based on the field distribution shown in Figs.~2(g) and 2(h), the two edge states with energy $E\approx \pm G$ in Figs.~\ref{FigS5}(f)-\ref{FigS5}(h) can be observed in the reflection spectra when the waveguide is coupled to the mechanical mode $b_{0}$.
Based on the above analyses, we can conclude that the \emph{edge states can be detected by the reflection spectra of a single phonon scattered in the waveguide}, which is coupled to a proper mechanical mode in the optomechanical ladder.

\section{adiabatic optomechanical pumping}

\begin{figure}[tbp]
\includegraphics[width=0.98 \textwidth]{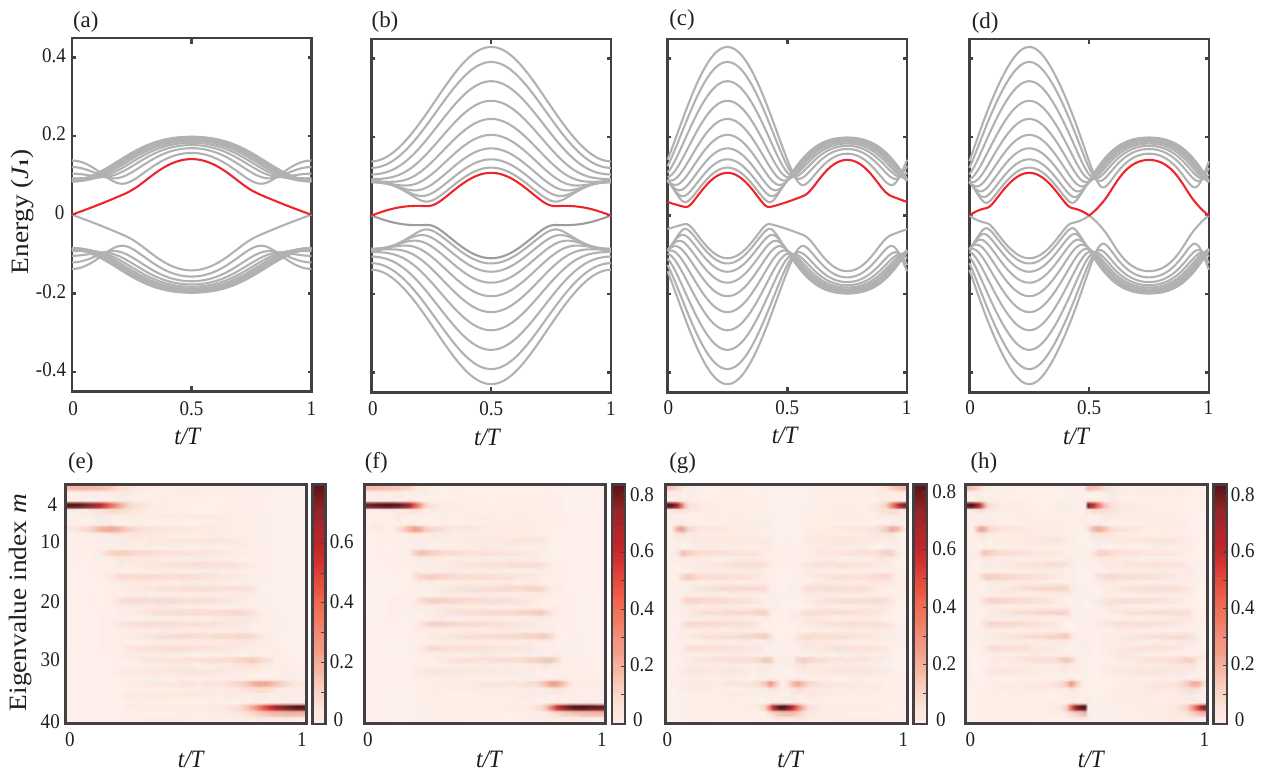}
\caption{(Color online) (a)-(d) Instantaneous energy spectra in an open chain ($N=10$) for four different modulation cases: (a) $\bar{G}=G_{c,1}$, $\tilde{G}=0.5(G_{c,2}-G_{c,1})$, $\tilde{\delta}=0.2$, $T=T^{\prime }$, $\phi_0=\phi'_0=0$, (b) $\bar{G}=G_{c,2}$, $\tilde{G}=-0.5(G_{c,2}-G_{c,1})$, $\tilde{\delta}=0.2$, $T=T^{\prime }$, $\phi_0=\phi'_0=0$, (c) $\bar{G}=0.5(G_{c,2}+G_{c,1})$, $\tilde{G}=-(G_{c,2}-G_{c,1})$, $\tilde{\delta}=0.2$, $T=T^{\prime }$, $\phi_0=\phi'_0=\pi/2$, and (d) $\bar{G}=0.5(G_{c,2}+G_{c,1})$, $\tilde{G}=-(G_{c,2}-G_{c,1})$, $\tilde{\delta}=0.2$, $T=2T^{\prime }$, $\phi_0=\phi'_0/2=\pi/2$. The red solid lines represent the evolutionary path of the edge states. (e)-(h) Time evolution of the probability distributions of the eigenstate corresponding to the red solid lines in the instantaneous energy spectra. Other parameters used are $J_{2}=0.6J_{1}$, $t_{1}=0.2J_{1}$, and $t_{1}=0.01J_{1}$.}
\label{FigS6}
\end{figure}

In this section, we show that the Chern numbers associated with the energy bands can also be verified by the adiabatic particle pumping processes.
The \emph{number of particles pumped} per cycle is an integer, which is given by a \emph{Chern number}~\cite{Janos2016Springer}.
In order to simulate a 2D Chern insulator in the optomechanical ladder, we replace the wave number for the second dimension by the time dimension in the periodical modulation of the driving frequency and strength.
Specifically, we introduce time-dependent optomechanical coupling strength $G(t)$ and detuning $\delta (t)$ as
\begin{equation}
G(t)=\bar{G}+\tilde{G}\cos (2\pi t/T + \phi_0),
\end{equation}%
and
\begin{equation}
\delta (t)=\tilde{\delta}\sin (2\pi t/T'+\phi'_0),
\end{equation}%
where $\bar{G}$, $\tilde{G}$, and $\tilde{\delta}$ are positive real numbers; $T$ ($T'$) is the modulation period and $\phi_0$ ($\phi'_0$) is the initial phase for the parameter $G(t)$ $[\delta (t)]$. Then the Hamiltonian of the \emph{periodically modulated optomechanical ladders} can be written as
\begin{eqnarray}
H_{\rm sys}(t) &=&\sum_{j}\left[ \delta (t)(a_{j}^{\dag }a_{j}-A_{j}^{\dag
}A_{j})-G(t)(a_{j}^{\dag }b_{j}+A_{j}^{\dag }B_{j}+\text{H.c.})\right] \notag\\
&&+\sum_{j}(J_{1}A_{j}^{\dag }a_{j}+J_{2}a_{j+1}^{\dag
}A_{j}+t_{1}B_{j}^{\dag }b_{j}+t_{2}b_{j+1}^{\dag }B_{j}+\text{H.c.}).
\end{eqnarray}%
We can investigate the adiabatic particle pumping processes based on the instantaneous energy spectra and the associated probability distributions.

\begin{figure}[tbp]
\includegraphics[width=0.98 \textwidth]{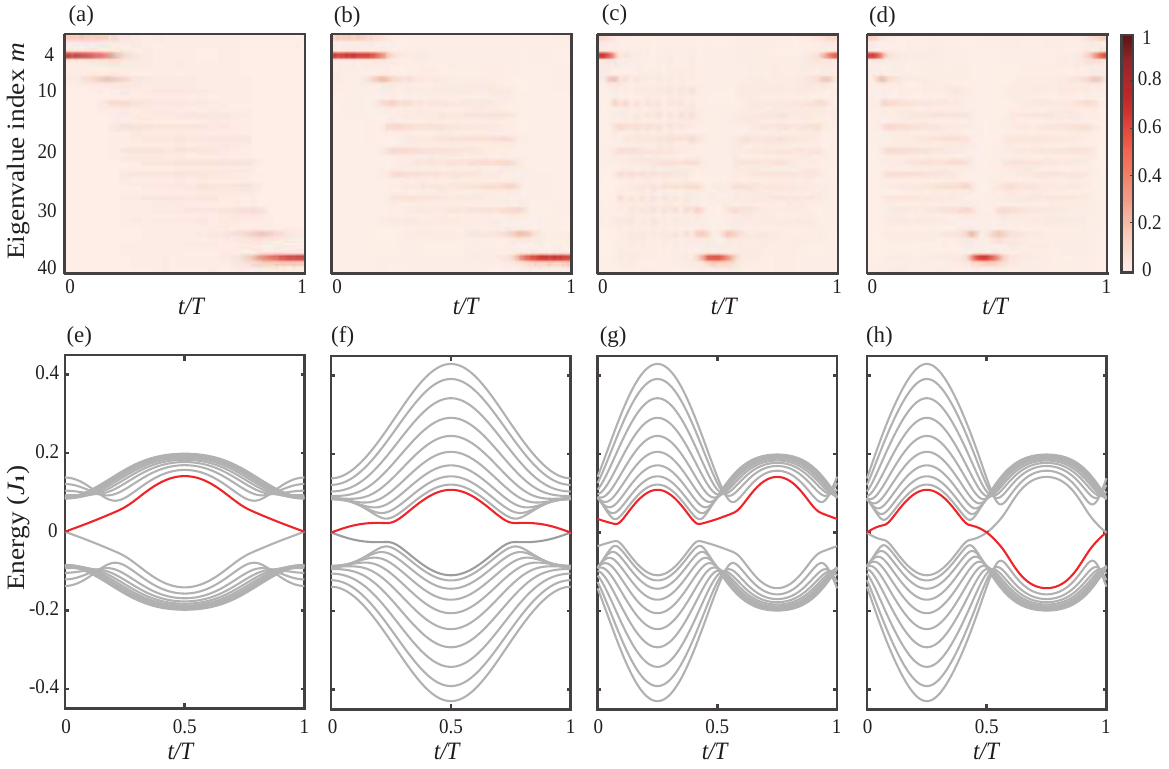}
\caption{(Color online) (a)-(d) Probability distribution for the adiabatic pump of the edge states in four different modulation cases with the same parameters used in Fig.~\ref{FigS4}. (e)-(h) Instantaneous energy spectra in an open optomechanical ladder with the red solid lines for the evolutionary path of the eigenstates.}
\label{FigS7}
\end{figure}

To better understand the \emph{adiabatic particle pumping processes}, we show the time evolution of the energy spectra (two of the four energy bands) by four different modulation schemes in Figs.~\ref{FigS6}(a)-\ref{FigS6}(d). In addition, the time evolution of the probability distributions for the eigenstate corresponding to the red solid lines in the instantaneous energy spectra are shown in Figs.~\ref{FigS6}(e)-\ref{FigS6}(h).

We assume that the probability of the eigenstate is initially localized around the mechanical mode $B_0$ in all the four cases.
In both the first and second cases [Figs.~\ref{FigS6}(e) and \ref{FigS6}(f)], the probability of the eigenstate pumps adiabatically to the right edge once in one period, corresponding to the Chern number $c_n=\pm 1$.
In the third case [Fig.~\ref{FigS6}(g)], the probability of the eigenstate moves from the left edge to the right edge and back to the left edge in one period,  corresponding to the Chern number $c_n=\pm 0$.
While in the fourth case [Fig.~\ref{FigS6}(h)], the probability of the eigenstate moves from the left edge to the right edge twice in one period, corresponding to the Chern number $c_n=\pm 2$.

Figures~\ref{FigS7}(a)-\ref{FigS7}(d) show the dynamics of the probability distribution for the adiabatic pumping of the edge states in different modulation cases, and the corresponding instantaneous energy spectra are shown in Figs.~\ref{FigS7}(e)-\ref{FigS7}(h).

Based on the probability distribution of the eigenstates shown in Figs.~\ref{FigS6}(e)-\ref{FigS6}(h), we choose the initial conditions with the probability distribution $P_{B_0}(0)=1$ and $P_{O}(0)=0$ for $O\neq B_0$.
Similar to Figs.~\ref{FigS6}(e)-\ref{FigS6}(g), the edge state is pumped adiabatically to the right edge once within a period in both the first and second cases [Figs.~\ref{FigS7}(a) and \ref{FigS7}(b)], and the edge state is pumped adiabatically to the right edge and then back to the left edge within a period in the third case [Fig.~\ref{FigS7}(c)].
Instead, Fig.~\ref{FigS6}(h) is quite different from Fig.~\ref{FigS7}(d) for the fourth case, and Fig.~\ref{FigS7}(d) is similar to the third case [Fig.~\ref{FigS7}(c)], i.e., the edge state is pumped adiabatically to the right edge and then back to the left edge within a period in the fourth case.
This means that both the third [Fig.~\ref{FigS7}(c)] and fourth [Fig.~\ref{FigS7}(d)] cases cannot be well distinguished by observing the adiabatic evolution process of the edge states.
However, it is worth mentioning that the adiabatic theorem is not applicable around the point ($t=0.5T$) due to the level crossing in instantaneous energy spectrum in Fig.~\ref{FigS7}(h).
From the instantaneous energy spectra, we can see that the energy of the solid red line for the third case is in the upper band while it is in the lower band for the fourth case during the time interval from $T/2$ to $T$, which provides an effective way to distinguish the third case from the fourth case.

We can confirm from the above discussions that the system can be extended to simulate a 2D Chern insulator by adiabatically modulating both the optomechanical strength and detuning. Moreover, we can observe the Chern numbers by the adiabatic particle pumping processes.

\end{document}